\documentclass[lettersize,journal]{IEEEtran}
\usepackage{amsmath,amsfonts}
\usepackage{algorithmic}
\usepackage{algorithm}

\usepackage{array}
\usepackage[caption=false,font=normalsize,labelfont=sf,textfont=sf]{subfig}
\usepackage{textcomp}
\usepackage{stfloats}
\usepackage{url}
\usepackage{verbatim}
\usepackage{graphicx}
\def\BibTeX{{\rm B\kern-.05em{\sc i\kern-.025em b}\kern-.08em
    T\kern-.1667em\lower.7ex\hbox{E}\kern-.125emX}}
\usepackage{balance}
\usepackage{color}
\usepackage{amssymb}
\usepackage{amsfonts}
\usepackage{amsmath}
\usepackage{citesort}

\usepackage{arydshln} 

    \def\Complex{{\rm\rule[.23ex]{.03em}{1.1ex}\kern-.3em{C}}}

    \newcommand{\be}{\begin{equation}} \newcommand{\ee}{\end{equation}}
    \newcommand{\bea}{\begin{eqnarray}} \newcommand{\eea}{\end{eqnarray}}
    \newcommand{\benum}{\begin{enumerate}} \newcommand{\eenum}{\end{enumerate}}

    \newcommand{\qa}{{\bf a}}

    \newcommand{\qk}{{\bf k}}

    \newcommand{\qp}{{\bf p}}
    
    \newcommand{\qr}{{\bf r}}
    \newcommand{\qs}{{\bf s}}
    
    \newcommand{\qu}{{\bf u}}
    \newcommand{\qv}{{\bf v}}
    \newcommand{\qw}{{\bf w}}

    \newcommand{\qA}{{\bf A}}
    \newcommand{\qB}{{\bf B}}

    \newcommand{\qF}{{\bf F}}
    
    \newcommand{\qH}{{\bf H}}
    \newcommand{\qI}{{\bf I}}

    \newcommand{\qR}{{\bf R}}
    
    \newcommand{\qT}{{\bf T}}
    \newcommand{\qU}{{\bf U}}

    \newcommand{\qY}{{\bf Y}}
    
    \newcommand{\qzero}{{\bf 0}}

    \newcommand{\qDelta}{{\boldsymbol \Delta}}
    
    \newcommand{\qpsi}{{\boldsymbol \psi}}

    \newcommand{\qalpha}{{\boldsymbol \alpha}}

    \newcommand{\qomega}{{\boldsymbol \omega}}
    \newcommand{\qphi}{{\boldsymbol\phi}}

    \newcommand{\bbR}{{\mathbb R}}
    \newcommand{\bbC}{{\mathbb C}}
    
    \newcommand{\bbU}{{\mathbb U}}

    \newcommand{\blkdiag}{{\sf blkdiag}}

\newtheorem{Theorem}{Theorem}

\begin{document}
\title{Beam Foreseeing in Millimeter-Wave Systems with Situational Awareness: Fundamental Limits via Cram\'{e}r-Rao Lower Bound}

\author{~Wan-Ting~Shih,~Chao-Kai~Wen,~\IEEEmembership{Fellow,~IEEE},~Shang-Ho~(Lawrence)~Tsai,~\IEEEmembership{Senior Member,~IEEE}, ~Shi~Jin,~\IEEEmembership{Fellow,~IEEE}, and~Chau~Yuen,~\IEEEmembership{Fellow,~IEEE}

    \thanks{{W.-T.~Shih} and {C.-K.~Wen} are with the Institute of Communications Engineering, National Sun Yat-sen University, Kaohsiung 80424, Taiwan, Email: {\rm  sydney2317076@gmail.com} and {\rm chaokai.wen@mail.nsysu.edu.tw}.}
    \thanks{{S.-H.~Tsai} is with the department of Electrical Engineering, National Yang Ming Chiao Tung University, Hsinchu 30010, Taiwan, Email: {\rm shanghot@mail.nctu.edu.tw}.}
    \thanks{{S.~Jin} is with the National Mobile Communications Research Laboratory, Southeast University, Nanjing 210096, P. R. China, Email: {\rm  jinshi@seu.edu.cn}.}
    \thanks{{C.~Yuen} is with the School of Electrical and Electronics Engineering, Nanyang Technological University, 639798 Singapore, Email: {\rm chau.yuen@ntu.edu.sg}.}
}

\maketitle

\begin{abstract}
Millimeter-wave (mmWave) networks offer the potential for high-speed data transfer and precise localization, leveraging large antenna arrays and extensive bandwidths. However, these networks are challenged by significant path loss and susceptibility to blockages. In this study, we delve into the use of situational awareness for beam prediction within the 5G NR beam management framework. We introduce an analytical framework based on the Cram\'{e}r-Rao Lower Bound, enabling the quantification of 6D position-related information of geometric reflectors. This includes both 3D locations and 3D orientation biases, facilitating accurate determinations of the beamforming gain achievable by each reflector or candidate beam. This framework empowers us to predict beam alignment performance at any given location in the environment, ensuring uninterrupted wireless access. Our analysis offers critical insights for choosing the most effective beam and antenna module strategies, particularly in scenarios where communication stability is threatened by blockages. Simulation results show that our approach closely approximates the performance of an ideal, Oracle-based solution within the existing 5G NR beam management system. 
\end{abstract}


\section{Introduction}
Millimeter-wave (mmWave) communication technology is pivotal in 5G and future mobile networks, offering substantial bandwidth and high throughput capabilities. However, its high-frequency spectrum encounters challenges due to harsh propagation conditions. Although the beamforming technique in mmWave devices provides energy focusing and mitigates losses, it is easily disrupted by obstacles and user equipment (UE) movement. To address this, the 3rd Generation Partnership Project (3GPP) 5G New Radio (NR) standard has implemented a beam management framework \cite{Rel15,tutorialNR}. Previous studies have extensively investigated the susceptibility of mmWave beams to blockages \cite{Block1,Block2,Block4,Block5,Block3,Block4,FastABS2,Zhang-21,Zhu-17TWC}.

In addition to offering high-throughput communication, mmWave communication also shows potential for high-accuracy localization, attributable to its large antenna arrays and wide bandwidths. Precise location information is beneficial for establishing mmWave links, which in turn enhances radio access network performance through increased throughput, reduced latency, and improved robustness. The synergy between mmWave communication and localization has been the subject of various studies \cite{ISAC-21Arxiv,JLC1,JLC2,JLC3,JLC4,JLC5,JLC6,JLC7,JLC8,JLC9,JLC10}. For instance, \cite{JLC4} explores simultaneous localization and channel estimation during pilot transmission. Similarly, \cite{JLC5} examines the impact of beamforming on localization accuracy and proposes strategies to minimize errors. Additionally, studies such as \cite{JLC6,JLC7,JLC8,JLC9,JLC10} highlight the importance of robust, location-aided beam alignment methods that account for both mobility and localization imperfections.

\begin{figure*}
    \begin{center}
        \resizebox{6.46in}{!}{%
            \includegraphics*{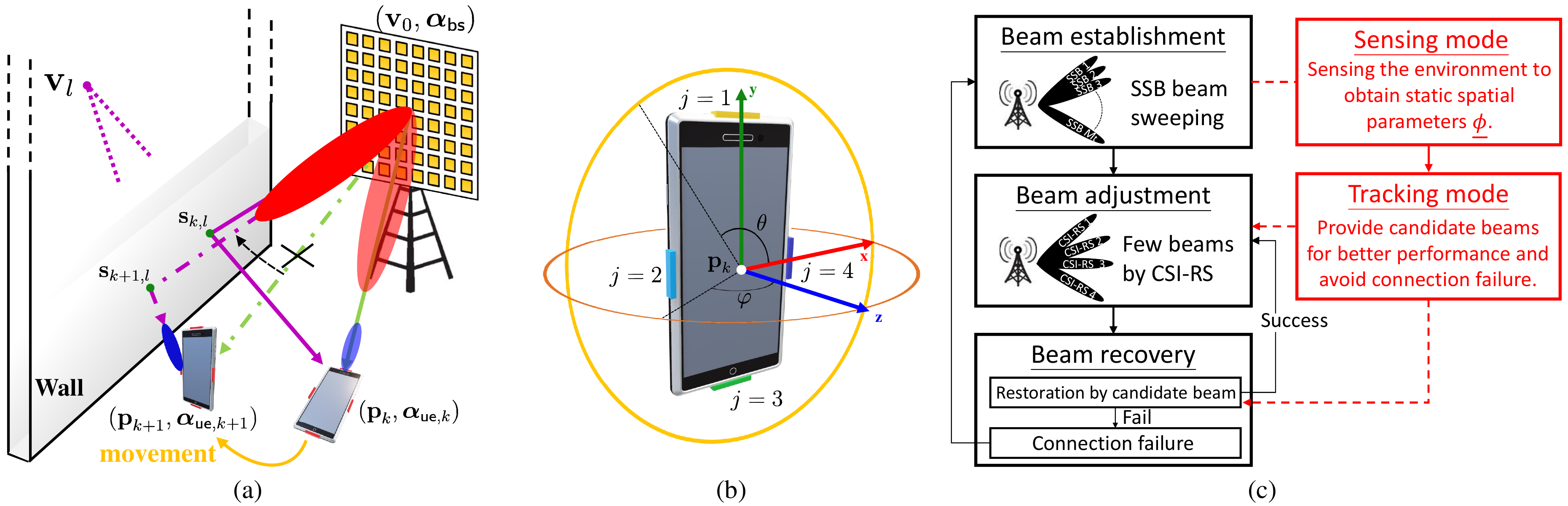} }%
        \caption{(a) Geometric model. (b) Placement of antenna panels on the mobile. (c) Block diagram of integration beam foreseeing into the 5G NR beam management.}\label{fig:awareness}
    \end{center}
\end{figure*}

While prior studies have mainly focused on location awareness in mmWave communication systems, our study delves into the potential of \emph{situational awareness} for beam prediction. Situational awareness, as defined in \cite{position2}, extends beyond location awareness to include knowledge of radio nodes and environmental landmark locations within their radio propagation. Fig. \ref{fig:awareness}(a) demonstrates how situational awareness can swiftly establish a secondary link during movement by considering the reflecting path. This path acts as a signal from a virtual base station (vBS), created by mirroring the physical BS. Unlike the changing reflection point, the vBS remains stationary, providing a stable reference for beam prediction. Our study also considers position-related information from multipath reflections for enhanced situational awareness.

Accurate landmark location information is vital for services based on situational awareness. Studies like \cite{mmWLoc1,mmWLoc2,mmWLoc3,mmWLoc5,position1,position3,position6,position7} emphasize precise position estimation of physical BSs, vBSs, and cluster nodes. However, a key challenge lies in balancing the limited wireless resources between communication and localization functions. The 5G NR introduces new positioning reference signals (PRS), designed specifically for positioning, enabling UEs to estimate localization-related information. While PRS assists in obtaining location information, attaining higher accuracy might require additional radio resources, leading to a trade-off between communication and localization \cite{JLC1,JLC2,JLC3}. In the context of commercially deployed mmWave 5G networks, several fundamental questions regarding the achievement of situational awareness and its impact on beam management remain:
\begin{itemize}
\item \emph{How is situational awareness implemented within the current 5G NR beam management framework?}
\item \emph{How does situational awareness contribute to beam management?}
\item \emph{How effective is the beam management approach?}
\end{itemize}

To answer these questions, this study introduces the concept of ``beam foreseeing''. It presents a seamless framework to align situational awareness with the current 5G NR beam management framework. Moreover, it proposes a novel method to analyze beamforming gains and predict beam alignment performance by evaluating the radio source from the physical BS and vBSs, where these environmental landmarks are obtained through situational awareness. This approach helps identify optimal beams for pairing and reduces the likelihood of connection failures, ensuring uninterrupted wireless access.
Our study makes the following contributions:
\begin{itemize}
\item We introduce a situational awareness procedure within the existing 5G NR beam management framework. Additionally, we develop a corresponding mathematical framework based on the Cram\'{e}r-Rao lower bound (CRLB) to quantify the position-related information from geometrical reflectors using received radio signals and visual-inertial odometry (VIO). This metric lays a crucial foundation for designing situational-aware services.

\item Our study analyzes the lower bound of the 6D static map of the environment, which includes 3D locations and 3D orientation biases. We evaluate the impact of channel quality and the number of received beams on these 6D estimates. We find that increasing the number of received beams has a limited effect on map information, while the map's accuracy is significantly influenced by the distance from the BS. Orientations are represented using quaternions to avoid issues associated with Euler angles. Moreover, we establish a physically meaningful lower bound on orientation bias in the form of quaternions, simplifying what is otherwise a complex task.

\item We demonstrate the relationship between position error bounds and beam foreseeing design. A robust beam alignment and antenna panel strategy are developed to ensure stable communication in scenarios with blockages. Simulation results indicate that the performance of our beam foreseeing approach closely aligns with that of the oracle solution in the current 5G NR beam management framework.
\end{itemize}

\emph{Notations}. $\mathbb{E}\{\cdot\}$ denotes the expectation operator, and $\| \cdot \|$ denotes the Euclidean norm. The notation $\qA\succeq \qB$ indicates that matrix $\qA-\qB$ is positive semidefinite. $\bbC^{m\times n}$ and $\bbR^{m\times n}$ represent sets of complex and real matrices of size $m\times n$, respectively. $\bbC^{m}$ and $\bbR^{m}$ represent sets of complex and real column vectors of size $m$, respectively. $\bbR^{m}\times\bbR^{n}$ indicates that the real column vector is composed of multiple components concatenated using the Cartesian product to represent their respective dimensions. The element in the $i$-th row and $j$-th column of matrix $\qA$ is represented as $[\qA]_{i, j}$ and the $i$-th element of vector $\qa$ is represented as $[\qa]_{i}$. Operator $\blkdiag(\cdot)$ represents block-diagonal matrices and $\mathrm{tr}\{\cdot\}$ is the trace of a square matrix. The superscripts $(\cdot)^\textrm{T}$ and $(\cdot)^\textrm{H}$ denote the transpose and conjugate transpose of a matrix, respectively. $\Re\{\cdot\}$, $\Im\{\cdot\}$, and $(\cdot)^*$ return the real part, imaginary part, and conjugate of a value, respectively.

\section{System Model and Problem Statement}
We consider a 5G mmWave MIMO system with a single BS using rectangular uniform planar array (R-UPA) with $N_{\sf t}$ transmit (Tx) patch antennas, a common BS array configuration in 5G. The UE is modeled with multiple panels, each consisting of $N_{\sf r}$ receive (Rx) patch antennas in a uniform linear array (ULA). The panels are placed along the edges of the UE with each side being indexed by $j$ (refer to Fig. \ref{fig:awareness}(b)), which ensures omnidirectional angular coverage of the UE. BS and UE use analog beamforming which is denoted by $\qw_{m_{\sf t}} \in \bbC^{N_{\sf t}}$ and $\qw_{m_{\sf r}} \in \bbC^{N_{\sf r}}$, respectively. The system is pre-assigned $M_{\sf t}$ and $M_{\sf r}$ beamforming vectors at the BS and UE.

\subsection{Geometric Model}
As shown in Fig. \ref{fig:awareness}(a), the UE is located at position ${\qp_k = [p_{{\sf x}, k}, \, p_{{\sf y}, k}, \, p_{{\sf z}, k}]^{\rm T}} \in \bbR^3$ with orientation $\qalpha_{{\sf ue},k}$ at time $k$.
The propagation paths from the BS (Tx) can reach the UE (Rx) through $L$ paths, where the first path ${(l=0)}$ is the line of sight (LoS) path, and the remaining paths $(l=1, \ldots, L-1)$ are the single reflection non-LoS (NLoS) paths.
The location and orientation of the BS are denoted by ${\qv_{0}  = [v_{{\sf x}, 0}, \, v_{{\sf y}, 0}, \, v_{{\sf z}, 0}]^{\rm T}} \in \bbR^3$ and $\qalpha_{{\sf bs}}$, respectively.
For each NLoS path, a reflection point $\qs_{k, l} \in \bbR^3$ exists on a reflecting surface. When the UE moves at time $k+1$, the reflection point $\qs_{k+1, l}$ moves. However, if we mirror the physical transmitter position on the reflecting surface, then we obtain the position $\qv_l=[v_{{\sf x}, l}, v_{{\sf y}, l}, v_{{\sf z}, l}]^{\rm T}\in \bbR^3$ of vBS which is static during the UE movement. Therefore, the radio channel from the accumulated times $K$ can be fully characterized by the {\bf spatial parameters},
\begin{multline}
\qphi = \Big[\underbrace{\qalpha_{{\sf ue},0}, \dots, \qalpha_{{\sf ue},K-1}, \qp_0^{\rm T}, \dots, \qp_{K-1}^{\rm T}}_{{\rm UE}},\\
\underbrace{~\qalpha_{{\sf bs}}, ~\qv_{0}^{\rm T}}_{\rm BS}, \underbrace{\qv_{1}^{\rm T}, \dots, \qv_{L-1}^{\rm T}}_{\rm vBS}  \Big]^{\rm T} \in \bbR^{7K} \times \bbR^4 \times \bbR^{3L}. \label{eq:phijk}
\end{multline}

\begin{figure*}[!b]
\vspace*{4pt}
\hrulefill
\begin{equation}
\qR(\qalpha) =
\begin{bmatrix}
\alpha_0^2+\alpha_1^2-\alpha_2^2-\alpha_3^2 & 2\alpha_1\alpha_2+2\alpha_0\alpha_3 & 2\alpha_1\alpha_3-2\alpha_0\alpha_2\\
2\alpha_1\alpha_2-2\alpha_0\alpha_3 & \alpha_0^2-\alpha_1^2+\alpha_2^2-\alpha_3^2 & 2\alpha_2\alpha_3+2\alpha_0\alpha_1\\
2\alpha_1\alpha_3+2\alpha_0\alpha_2 &
2\alpha_2\alpha_3-2\alpha_0\alpha_1 &
\alpha_0^2-\alpha_1^2-\alpha_2^2+\alpha_3^2
\end{bmatrix}\label{eq:Rotmatrix}
\end{equation}
\hrulefill
\begin{equation}\tag{4}\label{eq:geo}
\begin{aligned}
\tau_{k, l}&=\frac{\|\qv'_{l, k}\|}{c},~
\theta_{{\sf r}, k, l} = \tan^{-1}\left( \frac{v'_{{\sf y}, l, k}}{v'_{{\sf x}, l, k}}\right) ,~
\varphi_{{\sf r}, k, l} = \cos^{-1}\left( \frac{v'_{{\sf z}, l, k}}{\|\qv'_{l, k}\|}\right) ,\\
\theta_{{\sf t}, k, l}&=\left\lbrace \begin{array}{ll}
\tan^{-1}\left(\frac{p'_{{\sf y}, l, k}}{p'_{{\sf x}, l, k}}\right), & \text{if $\qv_{l}$ is not mirrored horizontally},\\
2\tan^{-1}\left(\frac{v'_{{\sf y}, l}}{v'_{{\sf x}, l}} \right) -\tan^{-1}\left(\frac{p'_{{\sf y}, l, k}}{p'_{{\sf x}, l, k}}\right)-\pi, & \text{if $\qv_{l}$ is mirrored horizontally}.
\end{array}\right.\\
\varphi_{{\sf t}, k, l}&=\left\lbrace \begin{array}{ll}
\cos^{-1}\left(\frac{p'_{{\sf z}, l, k}}{\|\qp'_{l, k}\|}\right), & \text{if $\qv_{l}$ is not mirrored vertically},\\
2\cos^{-1}\left(\frac{v'_{{\sf z}, l}}{\|\qv'_{l}\|} \right) -\cos^{-1}\left(\frac{p'_{{\sf z}, l, k}}{\|\qp'_{l, k}\|}\right)-\pi, & \text{if $\qv_{l}$ is mirrored vertically}.
\end{array}\right.
\end{aligned}
\end{equation}
\end{figure*}

In this study, we describe the positions of BS and UE in the global coordinate system (GCS). Their orientation describes the difference between the local coordinate system (LCS) and Global Positioning System, representing a rotation relative to the GCS. Various methods exist for representing orientation, including Euler angles, quaternions \cite{Ori}, and rotation matrices. Euler angles are intuitive and require only three angle values for sequential axis rotation but suffer from ambiguity and the Gimbal Lock problem\footnote{The relationship between the rotation state and its Euler Angles representation is not bijective, meaning that a single set of angles describes one rotation, but a rotation can have multiple sets of Euler Angles, leading to the ambiguity problem. The Gimbal Lock issue arises when one of the axes can no longer be rotated around due to the loss of one degree of freedom in the orientation.}. Comparing rotation matrices and quaternions, the latter can use fewer parameters to simplify the rotation. Consequently, we employ unit quaternions, which are four-dimensional complex numbers with a norm of one, i.e., $\qalpha = [\alpha_0, \alpha_1, \alpha_2, \alpha_3]$ with $\|\qalpha\|=1$.

The properties of unit quaternions offer several advantages, including the ability to multiply them together to form another unit quaternion and facilitate differential calculations. Unit quaternions can also be defined as a rotation around a unit orientation vector \cite{Ori}: ${\qalpha = [\cos(\theta/2), \sin(\theta/2)\qu^{\rm T} ]}$, where $\theta$ is the rotation angle and $\qu\in \bbR^3$ is the unit vector of the rotation axis. Hence, quaternions can represent a rotation of any angle $\theta$ from any orientation vector $\qu$ to another without the need to rotate according to a specific axis or order. For ease of derivation, unit quaternions can be converted into a rotation matrix, denoted as $\qR(\qalpha)$, as shown in \eqref{eq:Rotmatrix} below.

\subsection{Features of Channel Model}
We consider the geometry-based radio channel model \cite{WCbook}, as shown in Fig. \ref{fig:awareness}(a).
The spatial parameters are defined by the GCS, whereas the channel parameters, such as angle of arrival (AoA) and angle of departure (AoD), are usually observed through the LCS and thus depend on the orientations.
We transform the spatial parameters to the channel parameters via the {\bf geometric transformation}.
To describe the transformation, we first define the rotated position vectors:
\begin{subequations}
\begin{align}
\qv'_{l, k}&=[v'_{{\sf x}, l, k}, v'_{{\sf y}, l, k}, v'_{{\sf z}, l, k}] \triangleq\qR(\qalpha_{{\sf ue},k})^{-1}(\qv_{l}-\qp_k),\\
\qp'_{l, k}&=[p'_{{\sf x}, l, k}, p'_{{\sf y}, l, k}, p'_{{\sf z}, l, k}] \triangleq\qR(\qalpha_{{\sf bs}})^{-1}(\qp_k-\qv_{l}), \\
\qv'_{l}&=[v'_{{\sf x}, l}, v'_{{\sf y}, l}, v'_{{\sf z}, l}] \triangleq\qR(\qalpha_{{\sf bs}})^{-1}(\qv_{l}-\qv_{0}),
\end{align}
\end{subequations}
where the vectors on the left can be considered equivalent to the vectors on the right as shown by the LCS.
Then, the geometric transformation can be expressed as \eqref{eq:geo} below, where $\tau_{k, l}$, $(\theta_{{\sf r}, k, l}, \varphi_{{\sf r}, k, l})$, and $(\theta_{{\sf t}, k, l}, \varphi_{{\sf t}, k, l})$ denote the time of arrival (ToA), azimuth and elevation AoA, and azimuth and elevation AoD of the $l$-th path at time $k$, respectively; $c$ is the speed of light, and $\tan^{-1}(\cdot)$ denotes the four-quadrant inverse tangent.

As mentioned earlier, multiple antenna panels on UE are placed along its edges, as shown in Fig. \ref{fig:awareness}(b). We consider a far-field scenario, so the channel parameters for each antenna panel are similar. This condition is equivalent to assuming that the center of all antenna panels is at the center of the UE.
Then, the mmWave channel between the BS and the $j$-th antenna panel of UE at subcarrier $f_n \in\{ 0, \dots, N_s-1 \}$ and time $k$ can be expressed by an $N_{\sf r} \times N_{\sf t}$ matrix as
\begin{multline}\stepcounter{equation}\label{eq:channel}
\qH^{[j]}_k(f_n) =  \sqrt{N_{\sf t}N_{\sf r}} \Bigg(\sum_{l=0}^{L-1} e^{-j2\pi \frac{f_n}{N_s} \tau_{k, l}} \cdot g_{k, l}  \\
\cdot \qa_{{\sf r}}^{[j]}(\theta_{{\sf r}, k, l}, \varphi_{{\sf r}, k, l})\qa_{{\sf t}}^{\rm H}(\theta_{{\sf t}, k, l}, \varphi_{{\sf t}, k, l}) \Bigg),
\end{multline}
where $g_{k,l}$ is the corresponding complex-valued channel gain. In addition, $\qa_{{\sf t}}(\cdot)\in \bbC^{N_{\sf t}}$ and $\qa^{[j]}_{{\sf r}}(\cdot)\in \bbC^{N_{\sf r}}$ represent the array response vector of BS and each antenna panel of UE, respectively, and can be written as
$\qa_{{\sf t}}(\theta, \varphi) = \rho_{{\sf t}}(\theta, \varphi)\tilde{\qa}_{{\sf t}}(\theta, \varphi)$
and
$\qa_{{\sf r}}^{[j]}(\theta, \varphi) = \rho^{[j]}_{{\sf r}}(\theta, \varphi)\tilde{\qa}^{[j]}_{{\sf r}}(\theta, \varphi)$.
$\rho(\cdot)$ denotes the embedded antenna complex gain vector, and $\tilde{\qa}(\cdot)$ represents the ideal array response vector consisting of (hypothetical) isotropic antennas.
Taking the response vector of Rx as an example, the $m$-th element of $\tilde{\qa}_{{\sf r}}^{[j]}$ is denoted by
\begin{equation} \label{eq:idea_array}
[\tilde{\qa}^{[j]}_{{\sf r}}(\theta, \varphi)]_m = \frac{1}{\sqrt{N_{\sf r}}}\exp\left( -j\left( \qDelta^{[j]}_{{\sf r}, m}\right) ^{\rm T}\qk(\theta, \varphi)\right),
\end{equation}
where $\qDelta_{{\sf r}, m}$ is the relative position vector of the received antenna elements at each antenna panel with the origin at the center of each array, and $\qk(\theta, \varphi) \triangleq \frac{2\pi}{\lambda}[\cos\theta\sin\varphi, \, \sin\theta\sin\varphi, \, \cos\varphi]^{\rm T}$ is the wave vector with wavelength $\lambda$. We have a similar response vector for Tx.
We define the {\bf channel parameters} between the BS and the UE at time $k$ by the vector
\begin{equation}\label{eq:channel_par}
\qpsi_k = \begin{bmatrix}  \qpsi_{k, 0}^{\rm T}, \ldots,  \qpsi_{k, L-1}^{\rm T} \end{bmatrix}^{\rm T} \in \mathbb{R}^{7L},
\end{equation}
with
\begin{equation} 
\qpsi_{k, l} = \begin{bmatrix} \theta_{{\sf r}, k, l}, \,\varphi_{{\sf r}, k, l}, \,\theta_{{\sf t}, k, l}, \,\varphi_{{\sf t}, k, l},  \,\tau_{k, l}, \,g_{k, l}^{(\rm R)}, \,g_{k, l}^{(\rm I)} \end{bmatrix}^{\rm T}, \notag
\end{equation}
where $g_{k, l}^{(\rm R)}\triangleq\Re\{g_{k, l}\}$ and $g_{k, l}^{(\rm I)}\triangleq\Im\{g_{k, l}\}$ are the real and imaginary parts of $g_{k, l}$, respectively.

\subsection{Signal Model}
The transmission uses the orthogonal frequency-division multiplexing (OFDM) waveform according to 5G NR. Although the UE has multiple antenna panels that can receive signals, we limit our consideration to the use of one antenna panel at a time for signal reception, which is preferred because a single RF chain is a common architecture for early mmWave mobile phones.
The received signal for subcarrier $f_n$ at antenna panel $j$ at time (or sensing instance) $k$ is represented by $\qY^{[j]}_{k}(f_n)\in \bbC^{M'_{\sf r}\times M'_{\sf t}}$, where $M'_{\sf r} \leq M_{\sf r}$ and $M'_{\sf t} \leq M_{\sf t}$ are the number of used receive and transmit beams. We assume that the channel is time-invariant within $M'_{\sf r}\times M'_{\sf t}$ times because the movement of the user and changes in the environment are relatively slower than the millisecond-level beam management. As a result, the element of $\qY^{[j]}_{k}(f_n)$ from the $m_{\sf t}$-th Tx beam and $m_{\sf r}$-th Rx beam is given by
\begin{equation}\label{eq:rec_signal}
\left[ \qY^{[j]}_{k}(f_n)\right] _{m_{\sf r}, m_{\sf t}} = {\left(\qw_{m_{\sf r}}^{[j]}\right)^{\rm H}} \qH^{[j]}_k(f_n) \qw_{m_{\sf t}} + z^{[j]}_k(f_n),
\end{equation}
where $z^{[j]}_k(f_n)$ is the additive white Gaussian noise with variance $\sigma^2_{z}$.
In this work, we use RF phase shifters to implement the beamforming, and the beamforming vectors are expressed by
\begin{equation}\label{eq:weight}
\begin{aligned}
\qw_{m_{\sf t}} &=  \tilde{\qa}_{{\sf t}}(\theta_{m_{\sf t}}, \varphi_{m_{\sf t}}),~
\qw_{m_{\sf r}}^{[j]} = \tilde{\qa}_{{\sf r}}^{[j]}(\theta_{m_{\sf r}}^{[j]}, \pi/2),
\end{aligned}
\end{equation}
which, in fact, are designed on the basis of the ideal array response vector in \eqref{eq:idea_array}.
The beamforming vectors \eqref{eq:weight} focus transmit and received directions on $(\theta_{m_{\sf t}}, \varphi_{m_{\sf t}})$ and $\theta_{m_{\sf r}}^{[j]}$, respectively.
Notably, our analysis can also be applied to general beamforming vectors without restrictions on \eqref{eq:weight}.

\subsection{Beam Management}
In 5G NR \cite{Rel15,tutorialNR}, beam management comprises three stages: initial beam establishment \cite{IA1,IA2,IA3}, beam adjustment \cite{BT1,BT2}, and beam recovery, as illustrated by the black line in Fig. \ref{fig:awareness}(c). During initial beam establishment, a preliminary beam pair is established between a BS and a UE before data transmission. Specifically, the BS uses synchronization signal blocks (SSBs) in a time-multiplexed fashion, referred to as an SS burst set, to establish the beam pair link. The beamformed SSBs cover all directions in an SS burst set, with $M_{\sf t}$ beams transmitted periodically.

Beam adjustment involves updating the beam pair to maintain a robust connection, taking into account the UE's movements and rotations. The UE configures the downlink channel state information reference signals (CSI-RSs). Unlike the extensive beam-sweeping in initial beam establishment, the few beam directions result in a shorter measurement period, significantly less than $M_{\sf t}$.

Beam recovery is initiated when the connection between the BS and UE is lost due to significant movements or obstacles. Then, candidate beam pairs must be found to restore communication, with a beam rescanning being performed in cases of connection failure. Beam rescanning involves evaluating numerous beam combinations between the BS and UE's multiple antenna panels (e.g., $M_{\sf t} \times M_{\sf r} \times J$ beams), leading to unnecessary latency and signaling overhead.

\subsection{Problem Statement}
To assess beam alignment performance, we use the beamforming gain (BG) defined as
\begin{equation}\label{eq:BG}
\begin{aligned}
\textrm{BG}\left(\qw_{m_{\sf t}}, \qw_{m_{\sf r}}^{[j]} \right) &= \sum_{\forall f_n}\left|{\left(\qw_{m_{\sf r}}^{[j]}\right)^{\rm H}} \qH^{[j]}_k(f_n) \qw_{m_{\sf t}}\right|\\
&=\sum_{\forall f_n}\left|\tilde{H}^{[j]}_{k, m_{\sf r}, m_{\sf t}}(f_n)\right|,
\end{aligned}
\end{equation}
where $\tilde{H}^{[j]}_{k, m_{\sf r}, m_{\sf t}}(f_n) =  {\left(\qw_{m_{\sf r}}^{[j]}\right)^{\rm H}} \qH^{[j]}_k(f_n) \qw_{m_{\sf t}}$. By applying the beamformer in \eqref{eq:weight}, we obtain
\begin{multline}
\tilde{H}^{[j]}_{k, m_{\sf r}, m_{\sf t}}(f_n) =\sqrt{N_{\sf t}N_{\sf r}}\sum_{l = 0}^{L-1}g_{k, l}e^{-j2\pi \frac{f_n}{N_s}\tau_{k, l}}\cdot\rho^{[j]}_{{\sf r}}(\theta_{{\sf r}, k, l}, \varphi_{{\sf r}, k, l})\\
\cdot\underbrace{\left(\tilde{\qa}_{{\sf r}}^{[j]}(\theta_{m_{\sf r}}^{[j]}, \pi/2)\right) ^{\rm H}  \tilde{\qa}_{{\sf r}}^{[j]}(\theta_{{\sf r}, k, l}, \varphi_{{\sf r}, k, l})  }_{ \triangleq b^{[j]}_{{\sf r}, k, l}} \\
\cdot\rho_{{\sf t}}(\theta_{{\sf t}, k, l}, \varphi_{{\sf t}, k, l})\underbrace{\tilde{\qa}_{{\sf t}}^{\rm H}(\theta_{{\sf t}, k, l}, \varphi_{{\sf t}, k, l}) \tilde{\qa}_{{\sf t}}(\theta_{m_{\sf t}}, \varphi_{m_{\sf t}})  }_{\triangleq b_{{\sf t}, k, l}} . \label{eq:tildeH}
\end{multline}
The optimal beam pair is the one that achieves the highest BG for all possible values of $m_{\sf t} \in {1, \ldots, M_{\sf t} }$, $m_{\sf r} \in {1, \ldots, M_{\sf r} }$, and $j \in {1, \ldots, J }$.
To avoid unnecessary latency and signaling overhead in finding the optimal beam pair, we propose identifying suitable candidate beam pairs from spatial parameters as outlined in \eqref{eq:phijk}.

The objective of this paper is to evaluate the capability of 5G NR beam management in providing situational awareness and predicting the BG of candidate beams, as defined in \eqref{eq:BG}.
In Section III, we describe the methodologies employed to develop an efficient beam alignment scheme that leverages situational awareness derived from accumulated received signals. This section details the process of beam prediction and outlines strategies for beam and antenna panel selection to prevent scenarios that could lead to connection failures. Following this, Section IV discusses the use of the CRLB as a quantitative tool to assess and validate the effectiveness of our approach.

\section{Beam Foreseeing}

\subsection{Rationale}

Beam foreseeing uses the insight that, although the ideal beam pair changes with the UE's location and orientation, the locations of the transmit source, such as the BS and vBSs, remain fixed. This stability allows the prediction of the transmit source positions. Given the UE's state, we can identify candidate beam pairs by estimating the AoA and AoD between the UE and transmit sources.

\begin{figure}
    \begin{center}
        \resizebox{2.5in}{!}{%
            \includegraphics*{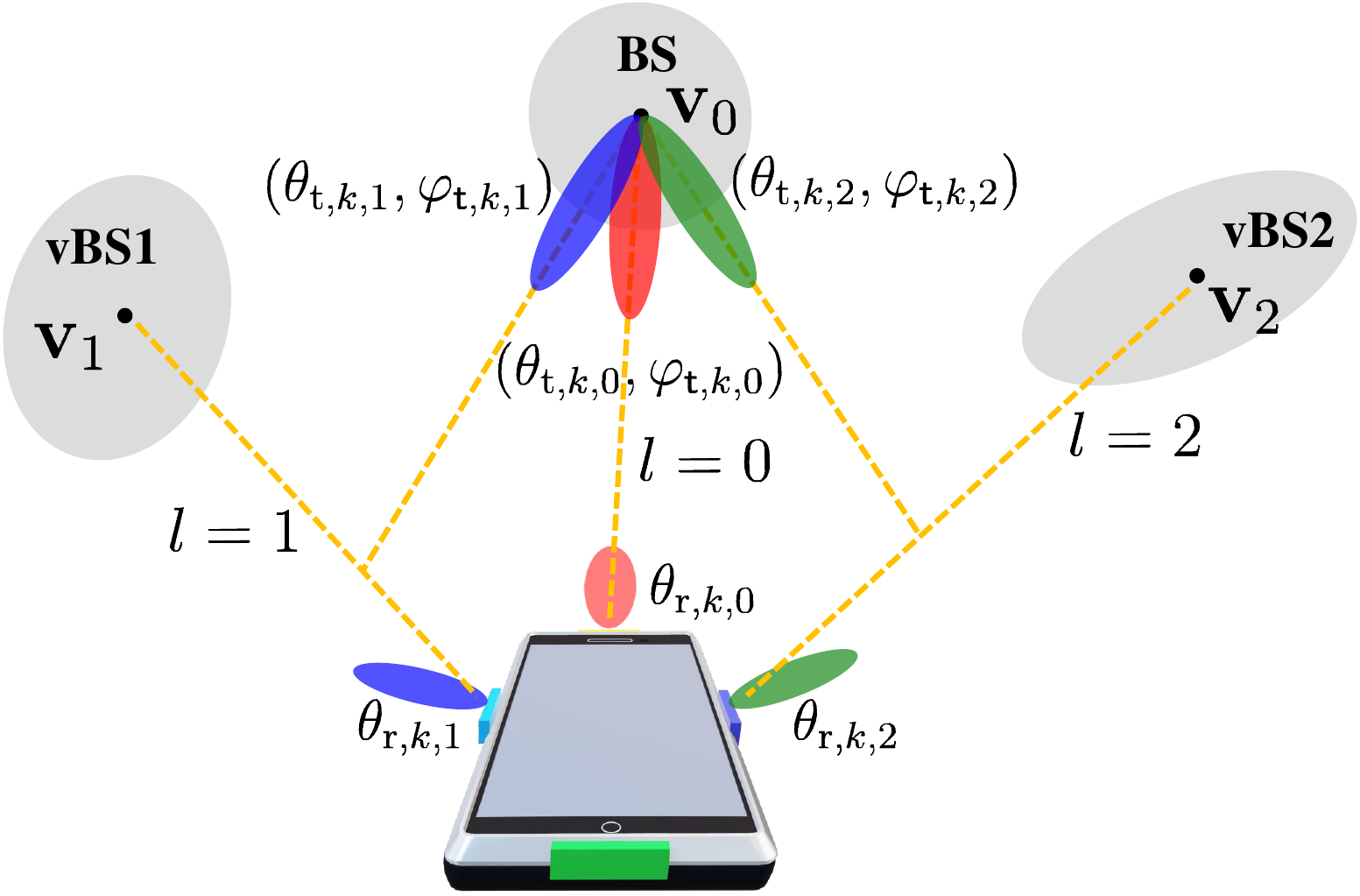}}%
        \caption{Rationale of beam foreseeing}\label{fig:Rationale}
    \end{center}
\end{figure}

Fig. \ref{fig:Rationale} shows the concept of beam foreseeing. Radio signals from the BS can reach the UE through three paths: the direct signal and two reflected signals from vBSs. We identify the presence of one BS and two vBSs by analyzing the environment. This scenario yields three possible beam pairs, namely, $(\theta_{m_{\sf t}}, \varphi_{m_{\sf t}}, \theta_{m_{\sf r}}^{[j]}) \in \{ (\theta_{{\sf t}, k, 0}, \varphi_{{\sf t}, k, 0}, \theta_{{\sf r}, k, 0}), (\theta_{{\sf t}, k, 1}, \varphi_{{\sf t}, k, 1}, \theta_{{\sf r}, k, 1})$, $(\theta_{{\sf t}, k, 2}, \varphi_{{\sf t}, k, 2}, \theta_{{\sf r}, k, 2}) \}$. This flexibility allows the beam pair to immediately switch to an alternate candidate in case of UE movement or obstruction, thereby maintaining uninterrupted communication.
Next, we explore the procedure of beam foreseeing in detail and propose a comprehensive switching strategy to select the optimal beam pair.

\subsection{Procedure}
As depicted in Fig. \ref{fig:awareness}(c), we integrate the concept of beam foreseeing into the existing 5G NR beam management procedure \cite{5GNR} through sensing and tracking modes:
\begin{itemize}
    \item {\bf Sensing Mode}: During initial beam establishment, we accumulate periodic and omnidirectional SS burst sets for environment sensing. This process includes estimating the locations and orientations of the BS and vBSs.
    
    \item {\bf Tracking Mode}: Building on the environmental landmarks identified in the sensing mode, the tracking mode efficiently generates candidate beam pairs. This approach aids in fine-tuning during beam adjustment and minimizes unnecessary latency from re-scanning during beam recovery. 
\end{itemize}

In the sensing mode, channel parameters obtained through SSBs are initially extracted using algorithms like NOMP and MUSIC \cite{NOMP,MUSIC}. We employ a subset of these channel parameters, as outlined in \eqref{eq:channel_par}, obtained through geometric transformation \eqref{eq:geo}, to determine the spatial parameters \eqref{eq:phijk}. Specifically, we exclude the AoDs from $\qpsi_{k, l}$ and use
\begin{equation} \label{eq:av_channelParms}
 \dot{\qpsi}_{k, l} = \begin{bmatrix}\theta_{{\sf r}, k, l}, \,\varphi_{{\sf r}, k, l},  \,\tau_{k, l}, \,g_{k, l}^{(\rm R)}, \,g_{k, l}^{(\rm I)} \end{bmatrix}^{\rm T}
\end{equation}
to denote the available channel parameters. Although channel parameters can also be obtained through CSI-RSs, their limited sensing capability primarily serves tracking the best beam pair rather than environment sensing. The performance of CSI-RS-assisted environment sensing will be discussed in Section V.B. 

Continuing, the environment information, such as the locations of the BS and reflecting surfaces, can be derived from available channel parameters using algorithms \cite{position1,position3}. As discussed in Section II.A, each reflecting surface generates a virtual source via a vBS. The locations of the physical BS and vBSs are fixed, enabling consistent environment characterization.

We set the initial UE position as the origin of the GCS (i.e., $\qp_0 = [0, 0, 0]^{\rm T}$) and define the LCS at the BS to ensure no rotation relative to the GCS. VIO provides the UE's trajectory, including relative positions and orientations at consecutive instances\footnote{VIO is the process of determining the relative position and orientation of a device in 3D space by analyzing associated camera images and inertial measurement unit data.}. Therefore, we assume that the UE's states $(\qp_k, \alpha_{{\sf ue}, k})$ and the BS orientation $\alpha_{{\sf bs}}$ are known and can be excluded from the unknown parameters as\footnote{We set the orientation of the BS as the reference and exclude it from the unknown parameters in $\underline{\qphi}$. The orientation of the UE at the initial time is transformed from $\alpha_{{\sf ue},0}$ to $\Delta\alpha$ in $\underline{\qphi}$. Alternatively, the orientation of the UE can also be used as a reference.}
\begin{equation}\label{eq:barphi}
  \underline{\qphi} = {\begin{bmatrix} \Delta\qalpha, \,\qv_{0}^{\rm T}, \,\qv_{1}^{\rm T}, \,\dots, \,\qv_{L-1}^{\rm T} \end{bmatrix}^{\rm T}} \in\bbR^4 \times \bbR^{3L},
\end{equation}
where $\Delta\qalpha^{\rm T} \in \bbR^4$ represents the orientation bias between the BS and the UE. Specifically, $\Delta\qalpha$ satisfies
\begin{equation}\label{eq:Qua_delta}
\qalpha_{\sf bs} = {\rm Q_m}(\Delta\qalpha, \qalpha_{{\sf ue},0}),
\end{equation}
where ${\rm Q_m}(\cdot)$ indicates quaternion multiplication \cite{Ori}.
Accuracy can be improved by estimating the static parameters $\underline{\qphi}$ over accumulated times or across multiple users. Notably, the number of static parameters for landmarks depends on the number of propagation paths, not the surfaces.

The propagation paths in the environment can be determined using $\underline{\qphi}$ and UE's location information.
Then, the estimated AoD $(\hat{\theta}_{{\sf t}, k, l}, \hat{\varphi}_{{\sf t}, k, l})$ and AoA $(\hat{\theta}_{{\sf r}, k, l}, \hat{\varphi}_{{\sf r}, k, l})$ pairs can be calculated via \eqref{eq:geo}.
Because analog beamforming\footnote{Fully digital or hybrid digital-analog beamforming architectures offer greater flexibility, scalability, and precision in acquiring the necessary channel parameters, thereby achieving enhanced situational awareness. Further research is necessary to evaluate the performance impact of estimation errors on these architectures, which is beyond the scope of this paper.} can only select a fixed number of beam directions, we select the beam directions with the closest angle expressed in the LCS relative to the orientation.
Given an estimated AoD $(\hat\theta_{{\sf t}, k, l}, \hat\varphi_{{\sf t}, k, l})$, we select the closest Tx beam $(\theta_{m_{\sf t}}, \varphi_{m_{\sf t}})$ from the codebook of $\theta_{m_{\sf t}}$ and $\varphi_{m_{\sf t}}$ based on $|\hat\theta_{{\sf t}, k, l}-\theta_{m_{\sf t}}|$ and $|\hat\varphi_{{\sf t}, k, l}-\varphi_{m_{\sf t}}|$, respectively. Similarly, we select the closest Rx beam $\theta_{m_{\sf r}}^{[j]}$ based on $|\hat{\theta}_{{\sf r}, k, l}-\theta_{m_{\sf r}}^{[j]}|$, where $\theta_{m_{\sf r}}^{[j]} = \theta_{m_{\sf r}}+\frac{\pi}{2}(j-1)$.

As discussed in Section II.D, only a limited number of candidate beam pairs are available using CSI-RS. Thus, the challenge lies in selecting candidate beams with reconnection opportunities from $L$ sets of estimated AoD and AoA pairs. We propose strategies to mathematically address the problem of selecting propagation paths and corresponding antenna panels in the subsequent subsection. The process of beam foreseeing is detailed in Algorithm \ref{alg:algBF}.

\begin{algorithm}[t]
\caption{Beam Foreseeing}\label{alg:algBF}
\begin{algorithmic}
\REQUIRE $K$: Accumulated times for positioning.
\ENSURE Candidate beam pairs.
\FOR{$k=1$ \TO $K$}
\STATE {\textsc{Sensing Mode}}
\STATE \hspace{0.5cm}\textbullet~ Estimate channel parameters $\dot{\qpsi}_{k, l}$ using \eqref{eq:av_channelParms}.
\STATE \hspace{0.5cm}\textbullet~ Update spatial parameters $\underline{\qphi}$ via \eqref{eq:barphi}.
\IF{confidence in transmit sources is high}
\STATE {\textsc{Tracking Mode}}
\STATE \textbullet~Construct $L$ sets of estimated AoD $(\hat{\theta}_{{\sf t}, k, l}, \hat{\varphi}_{{\sf t}, k, l})$ and AoA $(\hat{\theta}_{{\sf r}, k, l}, \hat{\varphi}_{{\sf r}, k, l})$ pairs using $\underline{\qphi}$ and \eqref{eq:geo}.
\STATE \textbullet~Select paths and antenna panels as per Algorithm \ref{alg:algSPA}.
\ENDIF
\ENDFOR
\end{algorithmic}
\end{algorithm}

\begin{figure*}[!b]
\vspace*{4pt}
\hrulefill
\begin{equation}\label{eq:opt_prob}\tag{17}
\begin{aligned}
\max_{l \in \{0,1,\ldots, L-1 \} \atop j \in \{1,2,\ldots, J \}}~~& N_s |g_{k, l}| |\rho^{[j]}_{{\sf r}}(\theta_{{\sf r}, k, l}, \varphi_{{\sf r}, k, l})| |\rho_{{\sf t}}(\theta_{{\sf t}, k, l}, \varphi_{{\sf t}, k, l})|\\
&\cdot \sqrt{N_{\sf t}N_{\sf r}}{\int^{\pi}_{-\pi}f_{\rm VM}(\hat{\theta}_{{\sf r}, k, l}) |b^{[j]}_{{\sf r}, k, l}|{\rm d }\hat{\theta}_{{\sf r}, k, l}}
 \cdot {\int^{\pi}_{0}\int^{\pi}_{-\pi}f_{\rm VM}(\hat{\theta}_{{\sf t}, k, l})f_{\rm VM}(\hat{\varphi}_{{\sf t}, k, l})|b_{{\sf t}, k, l}| {\rm d }\hat{\theta}_{{\sf t}, k, l}}{\rm d }\hat{\varphi}_{{\sf t}, k, l}
\end{aligned}
\end{equation}
\end{figure*}

\subsection{Selection of Paths and Antenna Panels}
As illustrated in Fig. \ref{fig:Rationale}, the estimated AoDs and AoAs contain errors due to inaccuracies in the transmit sources, depicted by the gray ellipse around the sources. These errors can lead to mismatched beam alignment.
To model the errors in estimated AoDs and AoAs, we utilize the von Mises distribution\footnote{The von Mises distribution holds substantial significance among distributions defined on the unit circle, akin to the Gaussian distribution's role in the linear continuum.}\cite{vonM}:
\begin{equation}\label{eq:vonM}
f_{\rm VM}(\hat{\theta}) = 1/\left( 2\pi I_0(\kappa)\right) e^{\kappa\cos(\hat{\theta}-\theta)}
\end{equation}
where $\kappa$ represents the inverse of the variance of the angle estimate, i.e., $\kappa = {1}/{\sigma^2_{\hat\theta}}$ and $I_0(\cdot)$ is the modified Bessel function of order zero. 
To determine which paths and antenna panels should be prioritized, we use the BG as defined in \eqref{eq:BG} as a metric for performance comparison. The optimal beam pair is the one that maximizes the BG, calculated based on the estimated AoDs and AoAs, while accounting for their errors. Specifically, when selecting the beam pair that aligns with the $l$-th path via the $j$-th antenna panel, we can predict the corresponding BG using
\begin{multline} \label{eq:BG_uncertain}
\int^{\pi}_{0}\int^{\pi}_{-\pi}\int^{\pi}_{-\pi}f_{\rm VM}(\hat{\theta}_{{\sf r}, k, l})f_{\rm VM}(\hat{\theta}_{{\sf t}, k, l})f_{\rm VM}(\hat{\varphi}_{{\sf t}, k, l})\\
\cdot\textrm{BG}\left(\qw_{m_{\sf t}}, \qw_{m_{\sf r}}^{[j]}\right) {\rm d }\hat{\theta}_{{\sf r}, k, l} {\rm d }\hat{\theta}_{{\sf t}, k, l} {\rm d }\hat{\varphi}_{{\sf t}, k, l}.
\end{multline}
To align with the estimated direction of the $l$-th path, we select the beam pair $(\qw_{m_{\sf t}}, \qw_{m_{\sf r}}^{[j]})$ closest to the estimated AoD $(\hat{\theta}_{{\sf t}, k, l}, \hat{\varphi}_{{\sf t}, k, l})$ and AoA $\hat{\theta}_{{\sf r}, k, l}$, considering the uncertainty of the angle estimations. This selection is equivalent to selecting the appropriate propagation path.
Using the Cauchy–Schwarz inequality, we can establish a lower bound for the beamforming gain on the $l$-th path in \eqref{eq:BG_uncertain}. This forms the basis of the optimization problem outlined in \eqref{eq:opt_prob} below. The terms in \eqref{eq:opt_prob} can be categorized into two groups:

\begin{figure*}[]
    \begin{center}
        \resizebox{5.7in}{!}{%
            \includegraphics*{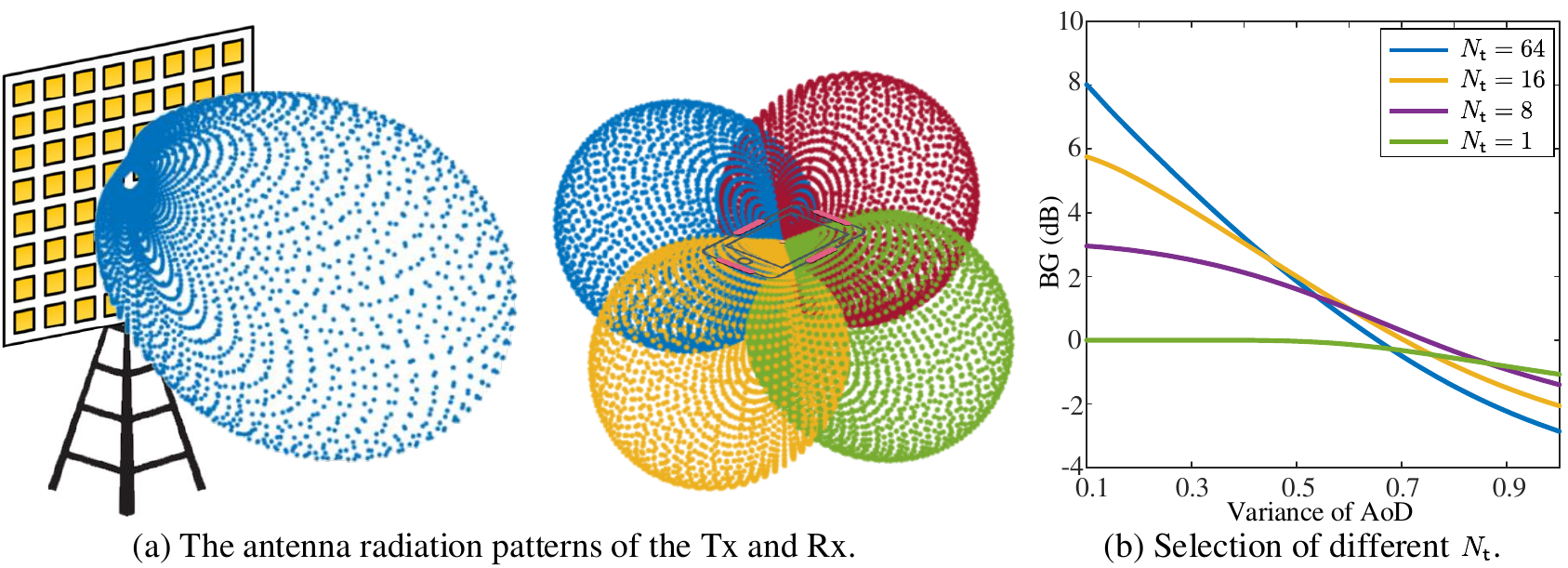}}%
        \caption{Comparison for selection of paths and antenna panels.}\label{fig:Selection}
    \end{center}
\end{figure*}

\begin{figure*}[!b]
\vspace*{4pt}
\hrulefill
\begin{equation}\label{eq:second}\tag{20}
\max_{l \in \{0,1,\ldots, L-1 \}}~~
\sqrt{N_{\sf r}}{\int^{\pi}_{-\pi}f_{\rm VM}(\hat{\theta}_{{\sf r}, k, l}) |b^{[j]}_{{\sf r}, k, l}|{\rm d }\hat{\theta}_{{\sf r}, k, l} } \cdot \sqrt{N_{\sf t}}{\int^{\pi}_{0}\int^{\pi}_{-\pi}f_{\rm VM}(\hat{\theta}_{{\sf t}, k, l})f_{\rm VM}(\hat{\varphi}_{{\sf t}, k, l})|b_{{\sf t}, k, l}| {\rm d }\hat{\theta}_{{\sf t}, k, l}}{\rm d }\hat{\varphi}_{{\sf t}, k, l}.
\end{equation}
\hrulefill
\begin{equation}\label{eq:Txbeam}\tag{21}
{\int^{\pi}_{0}\int^{\pi}_{-\pi}\sqrt{N_{\sf t}}f_{\rm VM}(\hat{\theta}_{{\sf t}, k, l})f_{\rm VM}(\hat{\varphi}_{{\sf t}, k, l}) \left| \left( \tilde{\qa}_{{\sf t}}(\theta_{m_{\sf t}}, \varphi_{m_{\sf t}})\right) ^{\rm H} \tilde{\qa}_{{\sf t}}(\theta_{{\sf t}, k, l}, \varphi_{{\sf t}, k, l})\right| {\rm d }\hat{\theta}_{{\sf t}, k, l} {\rm d }\hat{\varphi}_{{\sf t}, k, l}}.
\end{equation}
\end{figure*}

\begin{itemize}
\item {\bf Channel Impact}: Channel parameters such as channel gain and antenna gain are determined at the time of signal reception. The antenna gain, denoted by $\rho(\cdot)$, is influenced by the antenna's pattern and form factor. In mmWave communications, the performance of integrated antennas within a compact phone may be negatively affected by their proximity to other form factor elements, including cameras, glass displays, or microphones. Therefore, both path and antenna panel selection should take into account the channel and antenna gains. 
    
\item {\bf Beam Alignment Mismatch}: The environmental information collected might contain inaccuracies, leading to mismatched beam alignment. The limitations of a ULA with 1D beam-changing capabilities in a mobile device can further exacerbate this mismatch. Consequently, path selection should also account for this constraint.
\end{itemize}
Based on these observations, we can divide the optimization problem presented in \eqref{eq:opt_prob} into two distinct subproblems: 

The first subproblem relates to channel impact and is formulated as follows
\begin{equation}\stepcounter{equation}\label{eq:first}
\max_{l \in \{0,1,\ldots, L-1 \} \atop j \in \{1,2,\ldots, J \}}~~ |g_{k, l}| |\rho^{[j]}_{{\sf r}}(\theta_{{\sf r}, k, l}, \varphi_{{\sf r}, k, l})||\rho_{{\sf t}}(\theta_{{\sf t}, k, l}, \varphi_{{\sf t}, k, l})|.
\end{equation}
The objective is to select the optimal path and antenna panel that maximize channel and antenna gain. The key determinants for the value in \eqref{eq:first} are the path strength and the antenna radiation patterns of both the BS and the UE. We model the impact of the antenna radiation pattern on $\rho(\cdot)$ using the generalized antenna model from \cite{Rel14}. Fig. \ref{fig:Selection}(a) illustrates the Tx and Rx antenna radiation patterns, with parametrization as defined in \cite{antPat}. For a specific propagation path, the maximum antenna radiation pattern can be determined from the UE's four panels. The four antenna panels on the edge of the mobile device provide omnidirectional coverage, yielding similar values of $\max_{j \in \{1,2,\ldots, J \}}|\rho^{[j]}_{{\sf r}}(\theta_{{\sf r}, k, l}, \varphi_{{\sf r}, k, l})|$ for all $l=0, 1, \ldots, L-1$. Therefore, we can simplify problem \eqref{eq:first} to  
\begin{equation}
\max_{l \in \{0,1,\ldots, L-1 \}}~~ |g_{k, l}| |\rho_{{\sf t}}(\theta_{{\sf t}, k, l}, \varphi_{{\sf t}, k, l})|.
\end{equation}
Consequently, the two-dimensional search problem in \eqref{eq:first} reduces to selecting the path that maximizes channel and antenna gains, with the AoD constrained to not deviate significantly from the array's boresight. The final decision should be based on these gains.

The second subproblem of \eqref{eq:opt_prob} addresses beam alignment mismatch, as shown in \eqref{eq:second} below. We examine the error in Tx beam alignment, as expressed in \eqref{eq:Txbeam} below. To illustrate the impact of beam alignment, consider an example where each Tx/Rx antenna has infinite resolution, allowing for precise alignment of the estimated AoD/AoA, i.e., $(\theta_{m_{\sf t}}, \varphi_{m_{\sf t}}) = (\hat{\theta}_{{\sf t}, k, l}, \hat{\varphi}_{{\sf t}, k, l})$. Fig. \ref{fig:Selection}(b) depicts the BGs for different numbers of Tx antennas with fixed AoDs versus estimation variances. It demonstrates that the largest number of Tx antennas, which would typically provide the highest BG, is not necessarily optimal for larger variances, as a broader beamwidth may be more advantageous than a narrow one when the estimated AoD error is high. This characteristic applies similarly to errors caused by the Rx and is also relevant in scenarios with non-infinite resolution, such as with the coarse codebook we considered.

The error caused by the mobile phone, particularly the mismatch in elevation AoA,
\begin{equation*}
\left| \left( \tilde{\qa}_{{\sf r}}^{[j]}(\theta_{m_{\sf r}}^{[j]}, \pi/2)\right) ^{\rm H} \tilde{\qa}_{{\sf r}}^{[j]}(\theta_{{\sf r}, k, l}, \varphi_{{\sf r}, k, l})\right|,
\end{equation*}
is another significant factor influencing the BG. Therefore, when selecting the propagation path, the estimated elevation AoA $\hat{\varphi}_{{\sf r}, k, l}$ closest to $\pi/2$ should be chosen.

Communication through a LoS path between BS and UE is ideal in the absence of occlusions. However, this condition is not always achievable, especially considering various hardware limitations that may render NLoS paths more effective than LoS paths. In light of the preceding discussions, we propose the following strategies for beam and antenna panel selection: 1) Antenna panel selection should be based on the chosen path. 2) A beam with a broader beamwidth should be selected for its robustness to angle estimation errors. 3) Path selection should give priority to an AoD close to the R-UPA's boresight and an elevation AoA close to $\pi/2$ in the LCS on the mobile phone.
Based on these strategies, we outline the algorithm for the selection of paths and antenna panels in Algorithm \ref{alg:algSPA}.

\begin{algorithm}[]
\caption{Selection of Paths and Antenna Panels}\label{alg:algSPA}
\begin{algorithmic}
\REQUIRE $L$ sets of estimated azimuth and elevation AoDs $(\hat{\theta}_{{\sf t}, K, l}, \hat{\varphi}_{{\sf t}, K, l})$ and AoAs $(\hat{\theta}_{{\sf r}, K, l}, \hat{\varphi}_{{\sf r}, K, l})$.
\ENSURE Construction of candidate beam pairs $\big(\qw_{m_{\sf t}}, \qw_{m_{\sf r}}^{[j]} \big)$
\FOR{$l=0$ \TO $L-1$}
\STATE Add path $l$ to the candidate set if
\STATE \hspace{0.5cm}\textbullet~AoDs align with the R-UPA's boresight, and
\STATE \hspace{0.5cm}\textbullet~Elevation AoA is close to $\pi/2$.
\ENDFOR
\FOR{each candidate in the set}
\STATE \textbullet~Select the antenna panel based on the path.
\STATE \textbullet~Choose a beam with broad beamwidth for angle estimation error robustness.
\ENDFOR
\end{algorithmic}
\end{algorithm}

\section{Performance Analysis}
In beam foreseeing, candidate beam pairs are determined based on estimated AoDs and AoAs, which are derived from the estimation of spatial parameters $\underline{\qphi}$, as outlined in \eqref{eq:rec_signal}. This section focuses on determining the CRLB for $\underline{\qphi}$ and evaluating the performance of beam foreseeing using this derived CRLB.

\subsection{Localization Accuracy}
Let $\hat{\underline{\qphi}}$ denote the estimate of the spatial parameters. Note that as the orientation bias is a unit quaternion, the estimated $\Delta\hat{\qalpha}$ within $\hat{\underline{\qphi}}$ should be restricted to the set $ \bbU = \{ \qalpha: u(\qalpha) = 0 \}$ with $u(\qalpha) = \|\qalpha \|^2-1$. When the estimated vector $\hat{\underline{\qphi}}$ with $\Delta\hat{\qalpha}$ lying on the manifold $\bbU$, the mean squared error (MSE) of an estimator for $\underline{\qphi}$ is lower bounded by \cite{inequality,CCRB}:
\begin{equation}\stepcounter{equation}\stepcounter{equation}\label{eq:LB_def}
 \mathbb{E}\left\{\left( \hat{\underline{\qphi}}-\underline{\qphi} \right)^{\rm H} \left( \hat{\underline{\qphi}}-\underline{\qphi} \right) \right\}
 \succeq  \qU(\qU^{\rm T}\qF_{\underline{\qphi}}\qU)^{-1}\qU^{\rm T}.
\end{equation}
In the above equation, $\qF_{\underline{\qphi}}$ is the unconstrained Fisher information matrix (FIM) of $\underline{\qphi}$ with dimensions $(4+3L) \times (4+3L)$, and $\qU \in \bbR^{(4+3L) \times (3+3L)}$ is obtained by collecting the orthonormal basis vectors of null-space of the gradient vector $\partial u(\Delta\qalpha)/\partial\underline{\qphi} \in \bbR^{4+3L} $ with $\qU^{\rm T}\qU = \qI$.
The spatial parameters $\underline{\qphi}$ are obtained from the channel parameters $\qpsi_k$ through the geometric transformation \eqref{eq:geo}. Thus, we first focus on computing the FIM of the channel parameters.

The FIM of the channel parameters $\qpsi_k$ at time $k$ is given by the sum of the FIM for each subcarrier and beam pair in \eqref{eq:rec_signal}, which can be calculated element-wise as:
\begin{multline}\label{eq:channel_FIM}
\left[\qF_{\qpsi_k}\right]_{i, i'} = \\
\frac{2}{\sigma^2_{z}}\Re\left\{\sum_{m_{\sf r}=1}^{M'_{\sf r}}\sum_{m_{\sf t}=1}^{M'_{\sf t}}\sum_{f_n=0}^{N_s-1}\frac{\partial \left(\tilde{H}^{[j]}_{k, m_{\sf r}, m_{\sf t}}(f_n)\right)^{*}}{\partial[\qpsi_k]_{i}}\frac{\partial \tilde{H}^{[j]}_{k, m_{\sf r}, m_{\sf t}}(f_n)}{\partial[\qpsi_k]_{i'}}\right\}.
\end{multline}
Here, $\tilde{H}^{[j]}_{k, m_{\sf r}, m_{\sf t}}(f_n) $ is defined in \eqref{eq:tildeH}.
Appendix A provides the derivation of the elements of $\qF_{\qpsi^{[j]}_k}$.
Based on the derivation in Appendix A, the equivalent FIM (EFIM) for the available channel parameters can be obtained using the Schur complement as $\qF_{\dot{\qpsi}^{[j]}_k}$.

By applying the geometric transformation \eqref{eq:geo} based on the chain rule and the independent estimations from the $K$, the unconstrained FIM for the spatial parameters $\underline{\qphi}$ can be obtained as
\begin{equation}\label{eq:phi}
\qF_{\underline{\qphi}} = \sum^{K-1}_{k=0}\qT_{\dot{\qpsi}_k\rightarrow \underline{\qphi}}\qF_{\dot{\qpsi}_k} \qT_{\dot{\qpsi}_k\rightarrow \underline{\qphi}}^{\rm T},
\end{equation}
where $\qT_{\dot{\qpsi}_k\rightarrow \underline{\qphi}} = {\partial \dot{\qpsi}_k}/{\partial \underline{\qphi}^{\rm T}}$
is the transformation matrix. The matrix elements of $\qT_{\dot{\qpsi}_k\rightarrow \underline{\qphi}}$ are derived in Appendix B.
In particular, we decompose $\qT_{\dot{\qpsi}_k\rightarrow \underline{\qphi}}$ into
\begin{equation} \label{eq:T_psi}
\qT_{\dot{\qpsi}_k\rightarrow \underline{\qphi}} =
\left[
\begin{array}{ccc}
\qT_{\dot{\qpsi}_{k, 0}\rightarrow\Delta\qalpha} & \cdots & \qT_{\dot{\qpsi}_{k, L-1}\rightarrow\Delta\qalpha}\\ \hdashline[3pt/3pt]
\qT_{\dot{\qpsi}_{k, 0}\rightarrow\qv_{0}} &&\qzero\\
&\ddots&\\
\qzero&&\qT_{\dot{\qpsi}_{k, L-1}\rightarrow\qv_{L-1}}
\end{array}
\right].
\end{equation}
The channel realizations during the $K$ sensing instances should change, whereas the spatial parameters $\underline{\qphi}$ are time-invariant. This invariant characteristic is the main spirit of this work; thus, $\qF_{\underline{\qphi}}$ can be represented as a sum of information from the $K$ sensing instances.
We write
\begin{equation} \label{eq:qF_matrixUnfold}
	   \qF_{\underline{\qphi}}
    = \left[ \begin{array}{c;{3pt/3pt}ccc}
       \qF_{\Delta\qalpha} & \qF_{\qv_0, \Delta\qalpha}^{\rm T} & \cdots & \qF_{\qv_{L-1}, \Delta\qalpha}^{\rm T} \\ \hdashline[3pt/3pt]
       \qF_{\qv_0, \Delta\qalpha} &  \qF_{\qv_0}  & \cdots & \qzero \\
       \vdots &  \vdots  & \ddots & \vdots \\
       \qF_{\qv_{L-1}, \Delta\qalpha} &  \qzero  & \cdots & \qF_{\qv_{L-1}}
      \end{array}
      \right],
\end{equation}
where
\begin{subequations}
\begin{align}
\qF_{\Delta\qalpha} &= \sum^{K-1}_{k=0}\qT_{\dot{\qpsi}_{k, l}\rightarrow\Delta\qalpha}\qF_{\dot{\qpsi}_{k, l}}\qT_{\dot{\qpsi}_{k, l}\rightarrow\Delta\qalpha}^{\rm T}\in \bbR^{4 \times 4},\\
\qF_{\qv_l} &= \sum^{K-1}_{k=0}\qT_{\dot{\qpsi}_{k, l}\rightarrow \qv_l}\qF_{\dot{\qpsi}_{k, l}}\qT^{\rm T}_{\dot{\qpsi}_{k, l}\rightarrow\qv_l}\in \bbR^{3 \times 3},\label{eq:FIM_vl}\\
\qF_{\qv_l, \Delta\qalpha} &=\sum^{K-1}_{k=0}\qT_{\dot{\qpsi}_{k, l}\rightarrow\qv_l}\qF_{\dot{\qpsi}_{k, l}}\qT^{\rm T}_{\dot{\qpsi}_{k, l}\rightarrow\Delta\qalpha}\in \bbR^{3 \times 4},
\end{align}
\end{subequations}
and $\qF_{\dot{\qpsi}_{k, l}}$ is given by \eqref{eq:EFIM_feat}.
To account for the constrain of the unit quaternion, i.e., $ \Delta\qalpha \in \bbU$, the matrix $\qU$ can be represented as $\qU = \blkdiag(\qU_0, \qI_{3L})\in \bbR^{(4+3L) \times (3+3L)}$ \cite{CCRB}, where
\begin{equation}
\qU_0 = \left[
\begin{array}{ccc}
 \Delta\alpha_1/\Delta\alpha_0 & \Delta\alpha_2/\Delta\alpha_0 & \Delta\alpha_3/\Delta\alpha_0 \\ \hdashline[3pt/3pt]
 & -\qI_3&
\end{array} \right] \in \bbR^{4 \times 3}.
\end{equation}
Then, we obtain the constrained FIM $\tilde{\qF}_{\underline{\qphi}} = \qU^{\rm T}\qF_{\underline{\qphi}}\qU$ and define the other relevant matrices $\tilde{\qF}_{\Delta\qalpha} = \qU_0^{\rm T}\qF_{\Delta\qalpha}\qU_0$ and $\tilde{\qF}_{\qv_l, \Delta\qalpha} = \qF_{\qv_l, \Delta\qalpha}\qU_0$ for subsequent use.
Note that $\qF_{\qv_l}$ remains the same due to multiplied by the identity matrixes.

Using the EFIM analysis method, the constrained EFIM for $\qv_l$ can be expressed as:
\begin{equation}\label{eq:EFIM_vl}
    \tilde{\qF}_{\qv_l}^{\rm e} = \qF_{\qv_l}-\tilde{\qF}_{\qv_l, \Delta\qalpha} \cdot  \tilde{\qF}_{\Delta\qalpha}^{-1} \cdot \tilde{\qF}_{\qv_l, \Delta\qalpha}^{\rm T}.
\end{equation}
The EFIM in \eqref{eq:EFIM_vl} shows that the channel parameters $\dot{\qpsi}_{k, l'}$ for $l' \neq l$ do not affect the EFIM for the position of the $l$-th transmit source.
Therefore, when evaluating the localization accuracy of the $l$-th transmit source, we only need to focus on its channel parameters $\dot{\qpsi}_{k, l}$.
With the EFIM in \eqref{eq:EFIM_vl}, the position error bound (PEB) for transmit source $l$ can be calculated as follows:
\begin{equation}\label{eq:EB_dif}
	\textrm{EB}(\qv_l) = \sqrt{\mathrm{tr}\left\{ \left( \tilde{\qF}_{\qv_l}^{\rm e}\right) ^{-1} \right\}}.
\end{equation}
Through some derivations in Appendix C, we can obtain the orientation error bound (OEB) as
\begin{equation}\label{eq:EB_Dalpha}
	\textrm{EB}(\Delta\qalpha) = 2\cos^{-1}{\left(1-\frac{1}{2}\mathrm{tr}\left\{\left[ \qU ( \tilde{\qF}_{\underline{\qphi}}) ^{-1}\qU^{\rm T}\right] _{1:4, 1:4} \right\}\right) } .
\end{equation}
The error bound derived in \eqref{eq:EB_Dalpha} represents the physical angle error of the orientation vector, rather than the Euclidean distance of the quaternion, which is physically meaningful.


\subsection{Localization-based Angular Accuracy}

In Section III.C, we showed that the variance of the estimated AoDs and AoAs has a significant impact on BG. While we derived the PEB for the positions of the transmit sources previously, the BG depends on the angular directions of these sources, not their positions. Hence, to evaluate the BG, we must convert the location information of the UE and transmit sources into direction information. To differentiate the channel parameters estimated from the received signals and those estimated from the spatial parameters, we denote the estimated AoDs and AoAs as $(\ddot{\theta}_{{\sf t}, l}, \ddot{\varphi}_{{\sf t}, l})$ and $(\ddot{\theta}_{{\sf r}, l}, \ddot{\varphi}_{{\sf r}, l})$. The estimates are made after the $K$ sensing instances by using the location information of the UE and transmit sources. For simplicity, the time index superscript $K$ has been omitted.

The geometric transformation in \eqref{eq:geo} converts the spatial parameters $\underline{\qphi}$ into angle parameters. By using this transformation, we can calculate the angle error bound (AEB) for the azimuth and elevation AoD $(\ddot{\theta}_{{\sf t}, l}, \ddot{\varphi}_{{\sf t}, l})$ and AoA $(\ddot{\theta}_{{\sf r}, l}, \ddot{\varphi}_{{\sf r}, l})$ of each path, respectively.

\begin{Theorem}
	The AEB for the AoA $(\ddot{\theta}_{{\sf r}, l}, \ddot{\varphi}_{{\sf r}, l})$ and AoD $(\ddot{\theta}_{{\sf t}, l}, \ddot{\varphi}_{{\sf t}, l})$ from the spatial parameters are given by
	\begin{subequations}\label{eq:Theo2}
	\begin{align}
	\textrm{EB}(\ddot{\theta}_{{\sf r}, l}) &= \qomega_{\sf r, \theta}^{\rm T} \qF_{\qv, \Delta\qalpha}^{-1} \qomega_{\sf r, \theta}, ~~
	\textrm{EB}(\ddot{\varphi}_{{\sf r}, l}) = \qomega_{\sf r, \varphi}^{\rm T} \qF_{\qv, \Delta\qalpha}^{-1} \qomega_{\sf r, \varphi},\label{eq:Theo2ar}\\
	\textrm{EB}(\ddot{\theta}_{{\sf t}, l}) &= \qomega_{\sf t, \theta}^{\rm T} \qF_{\qv_0, \qv_l}^{-1} \qomega_{\sf t, \theta}, ~~
	\textrm{EB}(\ddot{\varphi}_{{\sf t}, l}) = \qomega_{\sf t, \varphi}^{\rm T} \qF_{\qv_0, \qv_l}^{-1} \qomega_{\sf t, \varphi},\label{eq:Theo2et}
	\end{align}
	\end{subequations}
    where
    \begin{equation} \label{eq:omega}
    \begin{aligned}
        \qomega_{\sf r, \theta}  &=  \begin{bmatrix} \frac{\partial \theta_{{\sf r}, l}}{\partial \Delta\qalpha} \frac{\partial \theta_{{\sf r}, l}}{\partial \qv_l^{\rm T}}\end{bmatrix}^{\rm T},~~
        \qomega_{\sf r, \varphi} =  \begin{bmatrix} \frac{\partial \varphi_{{\sf r}, l}}{\partial \Delta\qalpha} \frac{\partial \varphi_{{\sf r}, l}}{\partial \qv_l^{\rm T}}\end{bmatrix}^{\rm T},~~\\
        \qomega_{\sf t, \theta}  &=  \begin{bmatrix} \frac{\partial \theta_{{\sf t}, l}}{\partial \qv_0^{\rm T}} \frac{\partial \theta_{{\sf t}, l}}{\partial \qv_l^{\rm T}}\end{bmatrix}^{\rm T},~~
        \qomega_{\sf t, \varphi} =  \begin{bmatrix} \frac{\partial \varphi_{{\sf t}, l}}{\partial \qv_0^{\rm T}} \frac{\partial \varphi_{{\sf t}, l}}{\partial \qv_l^{\rm T}}\end{bmatrix}^{\rm T},
    \end{aligned}
	\end{equation}
	and $\qF_{\qv, \Delta\qalpha}$ and $\qF_{\qv_0, \qv_l}$ are defined in \eqref{eq:FvaFvv} below on this page.
\end{Theorem}
\begin{figure}
    \begin{center}
        \resizebox{2.3in}{!}{%
            \includegraphics*{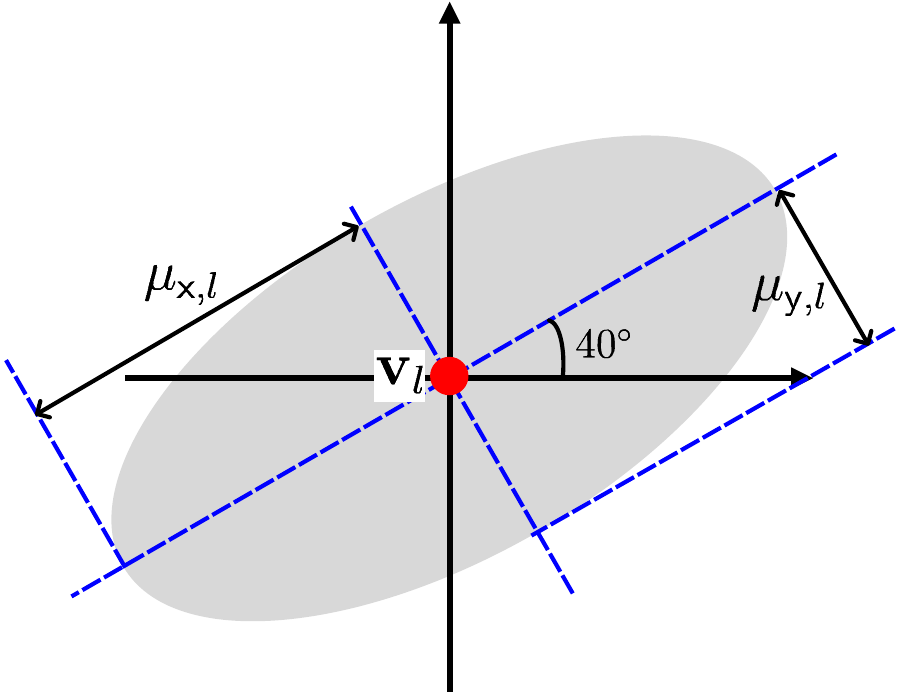}}%
        \caption{Information ellipse of $\qF_{\qv_l}$}\label{fig:ellipse}
    \end{center}
\end{figure}
\begin{figure*}[!b]
\vspace*{4pt}
\hrulefill
\begin{subequations}\label{eq:FvaFvv}
    \begin{align}
    \qF_{\qv, \Delta\qalpha} &= \begin{bmatrix} \qF_{\Delta\qalpha} - \sum_{l'\neq l}\qF_{\qv_{l'}, \Delta\qalpha}^{\rm T}\qF_{\qv_{l'}}^{-1}\qF_{\qv_{l'}, \Delta\qalpha} & \qF_{\qv_{l}, \Delta\qalpha}^{\rm T}\\
	\qF_{\qv_{l}, \Delta\qalpha}&\qF_{\qv_l}
	\end{bmatrix}, \label{eq:Fva} \\
	\qF_{\qv_0, \qv_l} &= \begin{bmatrix}\qF_{\qv_0}&0\\0&\qF_{\qv_l} \end{bmatrix}-
        \begin{bmatrix}
		\qF_{\qv_{0}, \Delta\qalpha} \\
		\qF_{\qv_{l}, \Delta\qalpha}
		\end{bmatrix}
		\left(\qF_{\Delta\qalpha} - \sum_{l'\neq l}\qF_{\qv_{l'}, \Delta\qalpha}^{\rm T}\qF_{\qv_{l'}}^{-1}\qF_{\qv_{l'}, \Delta\qalpha} \right) ^{-1}
        \begin{bmatrix}
		\qF_{\qv_{0}, \Delta\qalpha} \\
		\qF_{\qv_{l}, \Delta\qalpha}
		\end{bmatrix}^{\rm T}. \label{eq:Fvv}
	\end{align}
    \end{subequations}
\end{figure*}

\begin{proof}
The EB for $\ddot{\theta}_{{\sf r}, l}$ can be written as
	\begin{equation}\label{eq:EB_def}
	   \textrm{EB}(\ddot{\theta}_{{\sf r}, l})
    = \begin{bmatrix} \frac{\partial \theta_{{\sf r}, l}}{\partial \Delta\qalpha}& \frac{\partial \theta_{{\sf r}, l}}{\partial \qv_l^{\rm T}} &  \qzero \end{bmatrix}
      \qF^{-1}_{\underline{\qphi}}
      \begin{bmatrix} \frac{\partial \theta_{{\sf r}, l}}{\partial \Delta\qalpha}& \frac{\partial \theta_{{\sf r}, l}}{\partial \qv_l^{\rm T}} &  \qzero \end{bmatrix}^{\rm T}.
	\end{equation}
Using \eqref{eq:geo}, we find that the AoA of the $l$-th path is independent of the location of the $l'$ transmit source, resulting in ${\partial \theta_{{\sf r}, l}}/{\partial \qv_{l'}^{\rm T}} = 0$ for $ l' \neq l$. By inverting a $3\times 3$ block matrix and multiplying it with the vector in \eqref{eq:EB_def}, we obtain the non-zero block matrix of $\textrm{EB}(\ddot{\theta}_{{\sf r}, l})$ in \eqref{eq:Theo2ar}. Similar steps can be followed to compute EBs for $\ddot{\theta}_{{\sf t}, l}$, $\ddot{\varphi}_{{\sf r}, l}$, and $\ddot{\varphi}_{{\sf t}, l}$, with different differential terms.
\end{proof}
The following insights can be derived from Theorem 1.
\begin{itemize}
    \item \emph{Structure of FIMS}. \eqref{eq:FIM_vl} shows that the size of $\qF_{\qv_l}$ is $3\times3$, corresponding to the location of transmit source $l$. Thus, $\qF_{\qv, \Delta\qalpha}$ in \eqref{eq:Fva} has a block decomposition of EFIM for orientation bias $\Delta\qalpha$ and for the location of transmit source $l$, indicating the EB for $\ddot{\theta}_{{\sf r}, l}$ and $\ddot{\varphi}_{{\sf r}, l}$ depends on these parameters. Similarly, $\qF_{\qv_0, \qv_l}$ in \eqref{eq:Fvv} can be decomposed into the EFIM for BS ($l=0$) and vBS. Hence, the EB for $\ddot{\theta}_{{\sf t}, l}$ and $\ddot{\varphi}_{{\sf t}, l}$ depends on the estimates of BS and vBS.

     \item \emph{Known Orientation Bias}. In this case, the FIMs in \eqref{eq:Fva} and \eqref{eq:Fvv} simplify to diagonal matrices. The EB of AoA $(\ddot{\theta}_{{\sf r}, l}, \ddot{\varphi}_{{\sf r}, l})$ is only influenced by its corresponding transmit source, and the EB of AoD $(\ddot{\theta}_{{\sf t}, l}, \ddot{\varphi}_{{\sf t}, l})$ is only dependent on the BS and its corresponding vBS. The location information from other vBSs does not impact each other. We use the concept of the information ellipse \cite{theoretical} to gain deeper insights. Fig. \ref{fig:ellipse} shows an example of the information ellipse of $\qF_{\qv_l}$ in the xy-plane, with major and minor axes equal to $\mu_{{\sf x},l}$ and $\mu_{{\sf y},l}$. Assuming no orientation of UE, i.e., $\qR(\qalpha_{{\sf ue},K}) = \qI_{3\times 3}$, the EB of azimuth AoA $\ddot{\theta}_{{\sf r}, l}$ can be written as:
         \begin{equation}
            \textrm{EB}(\ddot{\theta}_{{\sf r}, l}) = \frac{\qu(\theta_{{\sf r}, l})^{\rm T} \qF_{\qv_l}^{-1} \qu(\theta_{{\sf r}, l}) }{ \left( v'_{{\sf x}, l, k}\right) ^2 + \left( v'_{{\sf y}, l, k}\right) ^2},
         \end{equation}
        where $\qu(\theta) = [-\sin(\theta)~\cos(\theta)]^{\rm T}$ is the unit vector, and $\qu(\theta_{{\sf r}, l})^{\rm T} \qF_{\qv_l}^{-1} \qu(\theta_{{\sf r}, l})$ is the Rayleigh quotient, which provides the observable value at direction $\theta_{{\sf r}, l}$. The Rayleigh quotient reaches its minimum value $\mu_{{\sf x},l}$ at directions $40^{\circ}$ and $220^{\circ}$, and its maximum value $\mu_{{\sf y},l}$ at directions $130^{\circ}$ and $310^{\circ}$. The information ellipse of $\qF_{\qv_l}$ heavily depends on the sensing directions and quality of the corresponding channel parameters. Thus, the sensing directions should cover all AoAs to achieve low EB for azimuth and elevation AoA 

        Notably, we do not measure the AoD of each path directly but instead use the AoAs from BS and vBSs to calculate $(\ddot{\theta}_{{\sf t}, l}, \ddot{\varphi}_{{\sf t}, l})$ through the geometric transformation \eqref{eq:geo}. Therefore, to obtain better AoD estimates, the sensing directions should have scanned the angle between the BS and vBS and the AoA from the transmit source. Moreover, the angle between the BS and vBS has twice the sensitivity to the EB for AoD $(\ddot{\theta}_{{\sf t}, l}, \ddot{\varphi}_{{\sf t}, l})$.

    \item \emph{Impact of Unknown Orientation}. When the orientation bias, $\Delta\qalpha$, is unknown and needs to be estimated, all estimates of the transmit sources are affected and lead to interrelated position estimation. The term $\qF_{\Delta\qalpha} - \sum_{l'\neq l}\qF_{\qv_{l'}, \Delta\qalpha}^{\rm T}\qF_{\qv_{l'}}^{-1}\qF_{\qv_{l'}, \Delta\qalpha}$ in $\qF_{\qv, \Delta\qalpha}$ and $\qF_{\qv_0, \qv_l}$ shows how the estimated performance of other paths affects that of the $l$-th path.
 \end{itemize}


\section{Simulation Results}

\begin{figure}
    \begin{center}
        \resizebox{2.9in}{!}{
                \includegraphics*{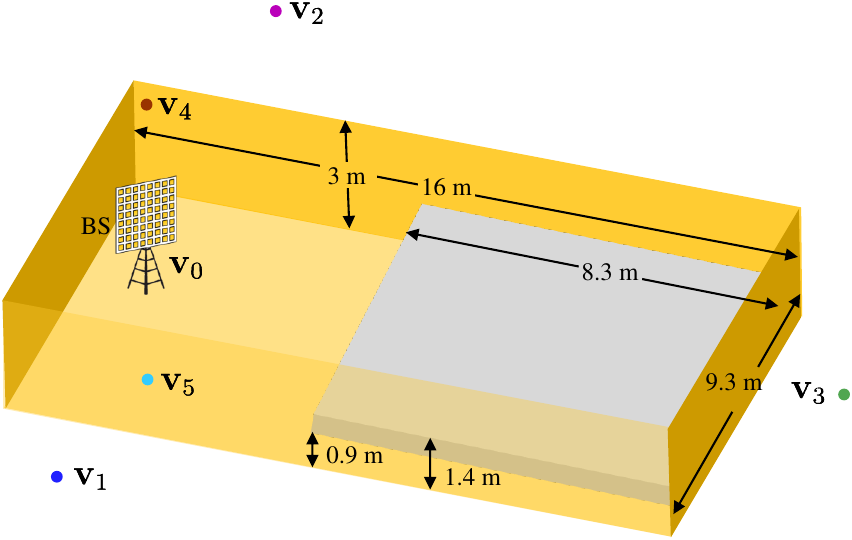}}
        \caption{Simulation scenario.}\label{fig:environment}
    \end{center}
\end{figure}

This section presents the findings of the previous analysis, considering the practical signal parameters. The scenario is an indoor environment with a 3\,m ceiling height and 30\,cm thick concrete walls, as illustrated in Fig. \ref{fig:environment}. One BS is deployed at a height of 2\,m and one UE is randomly deployed at any position with a random height between 0.9\,m and 1.4\,m and rotates randomly. The vBSs are determined through mirroring operations based on the BS position, as described in Section II.A, and denoted as $\qv_1$ to $\qv_5$, where $\qv_4$ and $\qv_5$ are mirrored by the ceiling and floor. The channel parameters are determined using the positions of vBSs and the wall segments, as outlined in \eqref{eq:geo}. The BS's R-UPA is unidirectional, indicating the beam cannot be transmitted from the back of the array. Therefore, the vBS mirrored by the wall facing away from the BS disappears. The complex channel gain $g_{k,l}$ depends on the reflective surface material and incidence angle, and the propagation paths are generated through ray tracing-based simulations using Wireless Insite$\circledR$ \cite{Wireless_Insite}. The UE has four antenna panels, as shown in Fig. \ref{fig:awareness}(b), and the antenna radiation patterns of the transmitter and receiver are shown in Fig. \ref{fig:Selection}(a).

We use OFDM signals with a carrier frequency $f_c = 28$ GHz, a bandwidth of 200\,MHz, and a subcarrier spacing of 60\,kHz. The BS and UE are equipped with R-UPA and ULA with $N_{\sf t} = 8\times 8$ and $N_{\sf r} = 4$. The codewords are set according to \eqref{eq:weight} with $(\theta_{m_{\sf t}}, \varphi_{m_{\sf t}}) \in \{ 20^{\circ}, 40^{\circ}, \ldots, 160^{\circ}\}$ and $M_{\sf t} = 64$ for the BS, and $\theta_{m_{\sf r}} \in \{30^{\circ}, 45^{\circ}, \ldots, 150^{\circ}\}$ and $M_{\sf r} = 9$ for the UE.
The performance is evaluated under three SNR conditions with different levels of noise variance: poor (best beam SNR below 5 dB), medium (best beam SNR between 5-15 dB), and good (best beam SNR beyond 15 dB).

\subsection{PEB, OEB, and AEB in Sensing Mode}

\begin{figure*}
    \begin{center}
        \resizebox{6.3in}{!}{%
            \includegraphics*{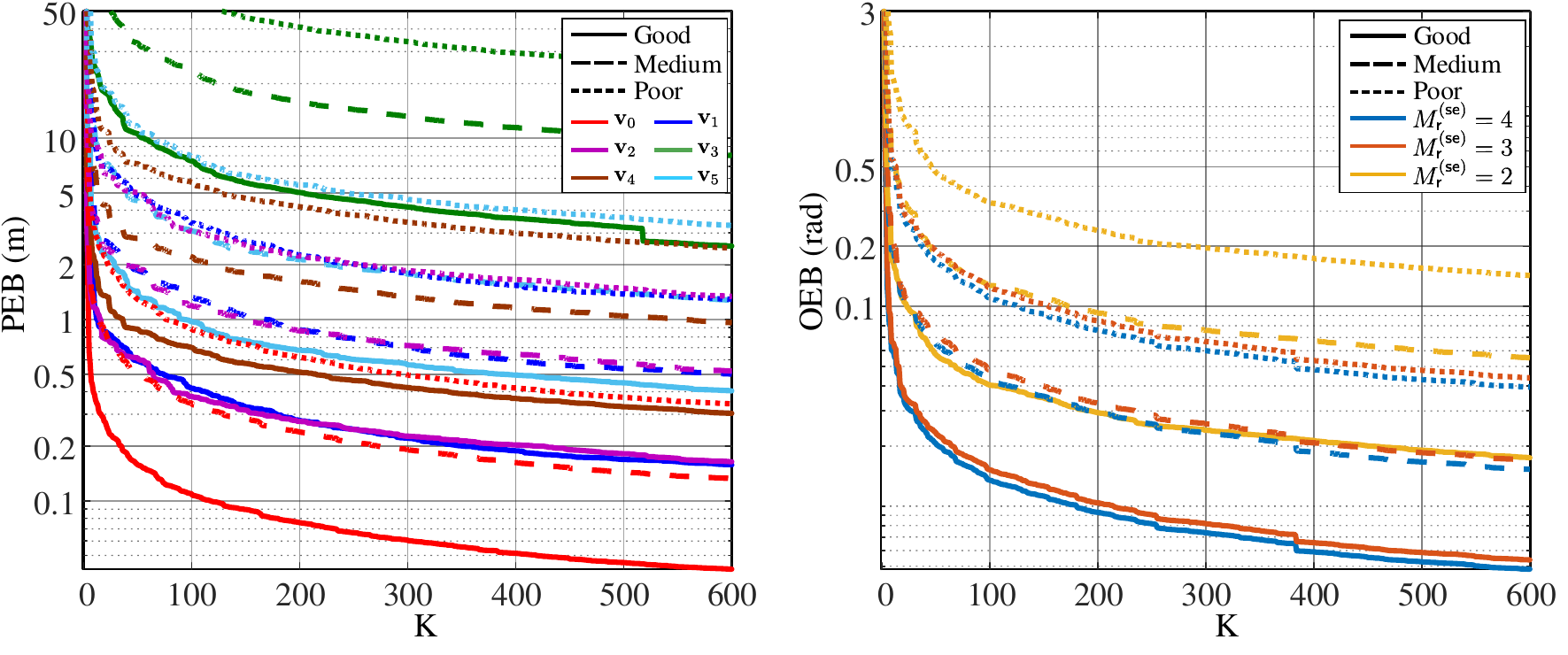}}%
        \caption{(a) PEB as a function of the accumulated time $K$ under different SNR conditions and (b) OEB as a function of the accumulated time $K$ under different SNR conditions and different numbers of Rx beams.}\label{fig:PEB}
    \end{center}
\end{figure*}

\begin{figure*}
    \begin{center}
        \resizebox{6.6in}{!}{%
            \includegraphics*{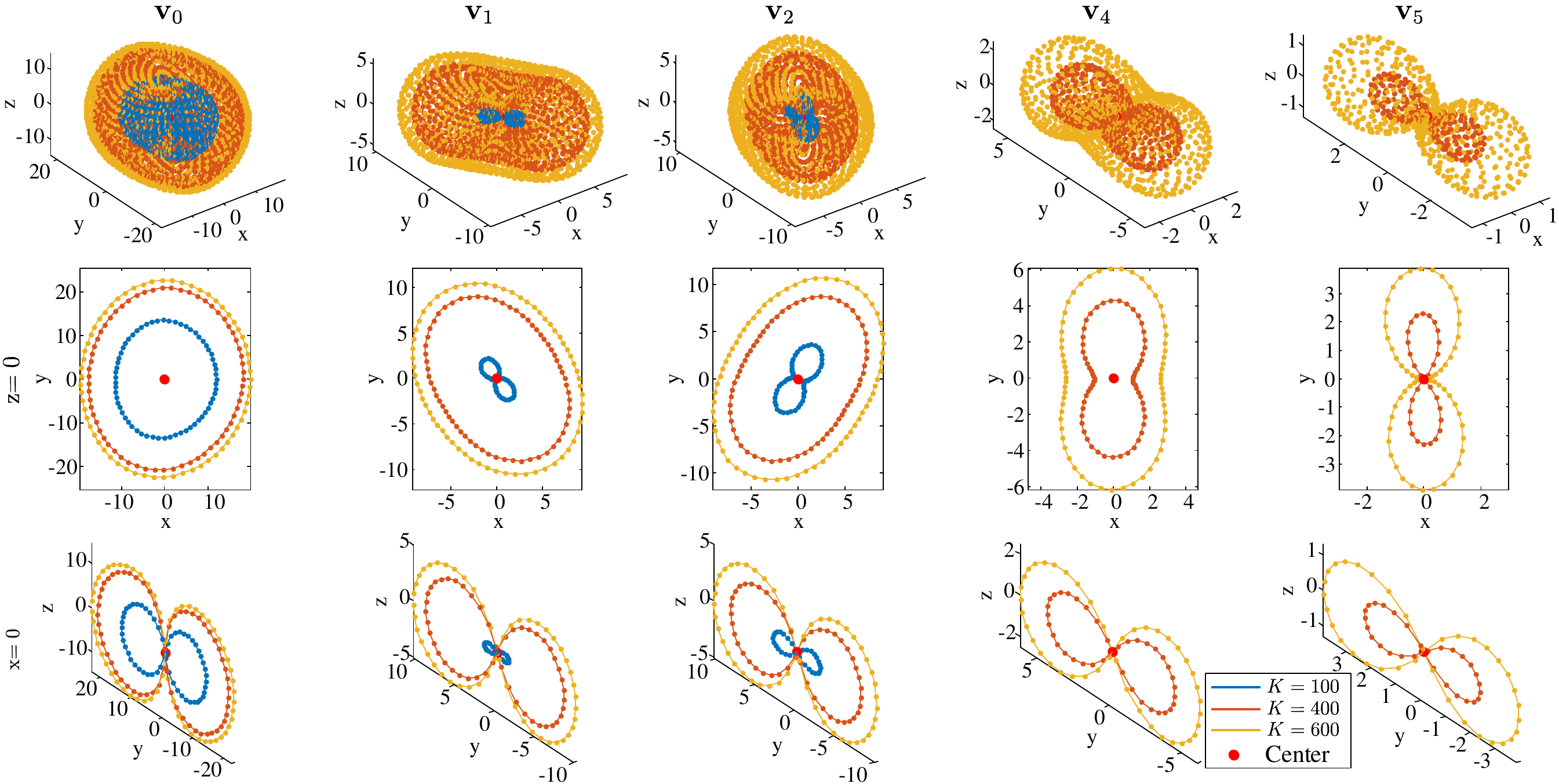}}%
        \caption{AEB in dB plotted against different angles for AoA at different positions of transmit source.}\label{fig:AEB}
    \end{center}
\end{figure*}
In this subsection, we aim to examine the impact of various SNR conditions, the number of Rx beams used for channel estimation, and the number of sensing instances on the positioning accuracy for different transmit sources and the orientation bias accuracy in the sensing mode. We evaluate the positioning accuracy of each transmit source using the PEB, as given in \eqref{eq:EB_dif}. Additionally, we assess the orientation bias accuracy using the OEB, as defined in \eqref{eq:EB_Dalpha}. In the sensing mode, the channel measurements are based on the SSBs with $N_s^{{\sf SSB}} = 240$ subcarriers and $M_{\sf t}^{(\sf se)} = M_{\sf t} = 64$ transmit beams.
The UE uses power detectors to determine the antenna panel and the corresponding Rx beam for channel estimation. The specific angles used for channel estimation are $\{45^{\circ}, 75^{\circ}, 105^{\circ}, 135^{\circ}\}$, $\{60^{\circ}, 90^{\circ}, 120^{\circ}\}$, or $\{60^{\circ}, 90^{\circ}\}$ for four-beam ($M_{\sf r}^{(\sf se)} = 4$), three-beam ($M_{\sf r}^{(\sf se)} = 3$), and two-beam ($M_{\sf r}^{(\sf se)} = 2$), respectively.
Note that the UE can only be deployed in the gray area depicted in Fig. \ref{fig:environment}.

Fig. \ref{fig:PEB}(a) shows the PEB under different SNR conditions with $M_{\sf r}^{(\sf se)} = 4$. The horizontal axis represents sensing instances $K$, while the different colored lines represent different transmit sources. Fig. \ref{fig:PEB}(a) reveals as the SNR condition deteriorates, the number of sensing instances required to achieve a desired PEB increases, and the final PEB is larger than that in other SNR conditions.
In addition, the PEB results of $\qv_1$ and $\qv_2$ are similar in Fig. \ref{fig:PEB}(a) because the positions of $\qv_1$ and $\qv_2$ are symmetrical with respect to the BS for the positions of UE. Furthermore, the PEB results for different vBSs are presented in relation to the distance to the BS. The farthest from the BS, $\qv_3$, has the worst result. The BS is located closer to the ceiling; thus, the PEB of the reflection source $\qv_4$ formed by the ceiling is better than that of the reflection source $\qv_5$ formed by the floor.
Finally, the stair-like results for $\qv_3$ are due to the fact that the information of $\qv_3$ is unavailable at certain sensing instances. The UE would rotate randomly in the gray area of Fig. \ref{fig:environment} at different sensing instances. However, the UE can only power one antenna panel to receive the signal at a time. As a result, the incident signal from $\qv_3$ is behind the boresight of the used antenna panel and thus cannot be received.

Fig. \ref{fig:PEB}(b) displays the OEB under different SNR conditions and with varying numbers of Rx beams. Unlike PEB, OEB presents the combined result of transmit sources, making it a more comprehensive indicator of performance under different conditions. Fig. \ref{fig:PEB}(b) indicates three or four Rx beams have a negligible effect on OEB, while a significant degradation is observed when reducing the beam number to two. Specifically, given an OEB target, the number of Rx beams that need to be scanned for each sensing instance in the sensing mode should be determined based on the communication quality. For example, in the medium SNR condition, an OEB of 0.2\,rad can be achieved after $K=50$ even when using only two Rx beams. Therefore, it is unnecessary to scan all Rx beams in one sensing instance, as the FIM for the position of the transmit source is incrementally improved through different positions of the UE at different sensing instances, leading to a reduction in PEB.

In Section IV.B, we explain that the BG is directly related to the directions of the transmit sources rather than their positions. Therefore, we investigate the AEB of every direction for different transmit sources. Under medium SNR and $M_{\sf r}^{(\sf se)} = 4$, Fig. \ref{fig:AEB} shows the AEB for each transmit source in every direction with different $K$. The red dot in the figure represents the reference point of the polar coordinate, and the radius in each angle is given by $-10 \log_{10}\textrm{EB}(\ddot{\theta}_{{\sf r}, l})$. The contour represents the observable accuracy at each angle. Notably, we set the radius as zero when the calculated radius is less than zero, indicating that the angle does not provide any useful information if the AEB is greater than $1$. For ease of observation, we also display the results of profile ${\rm z}=0$ and ${\rm x}=0$.

Fig. \ref{fig:AEB} shows the shape of the AEB maintained while gradually expanding outward with the increase in $K$. The shape is determined by the location of the transmit sources. For example, the AEB contour for $\qv_0$ appears similar to a ball, indicating the angular information with respect to the BS has nearly equal accuracy at every angle. This approach is reasonable because the BS is placed in the center of the environment, and the UE can acquire various angular information from the BS. However, the AEBs for $\qv_1$ and $\qv_2$ have a certain angular preference, and their AEB contours in ${\rm z}=0$ are symmetric to each other because the UE can only acquire certain angle information from the vBSs. Moreover, the AEBs for $\qv_4$ and $\qv_5$ have similar contours, but $\qv_4$ is better. The disappearance of $\qv_3$ indicates the AEB of $\qv_3$ is not less than zero. Thus, $\qv_3$ cannot obtain a sufficiently accurate angle. The AEB contours of the AoD are similar to that of the AoA, suggesting the estimation quality of AoA and AoD regarding a transmit source is similar. As mentioned in Section IV.B, the AEB contour for the AoD is the combination of the AEB contours from the AoA of BS and vBSs.

In summary, all error bounds improve with an increase in the accumulation times, the SNR condition, and the number of Rx beams. The choice of the number of Rx beams depends on the current SNR condition and the accumulated time. With sufficient accumulation time and the medium SNR condition, only a few Rx beams are required to estimate AoA. Moreover, the error bounds of virtual transmit sources are affected by their distance relative to the BS.

\subsection{Performance with CSI-RS-aided}
\begin{figure}
    \begin{center}
        \resizebox{3.5in}{!}{%
            \includegraphics*{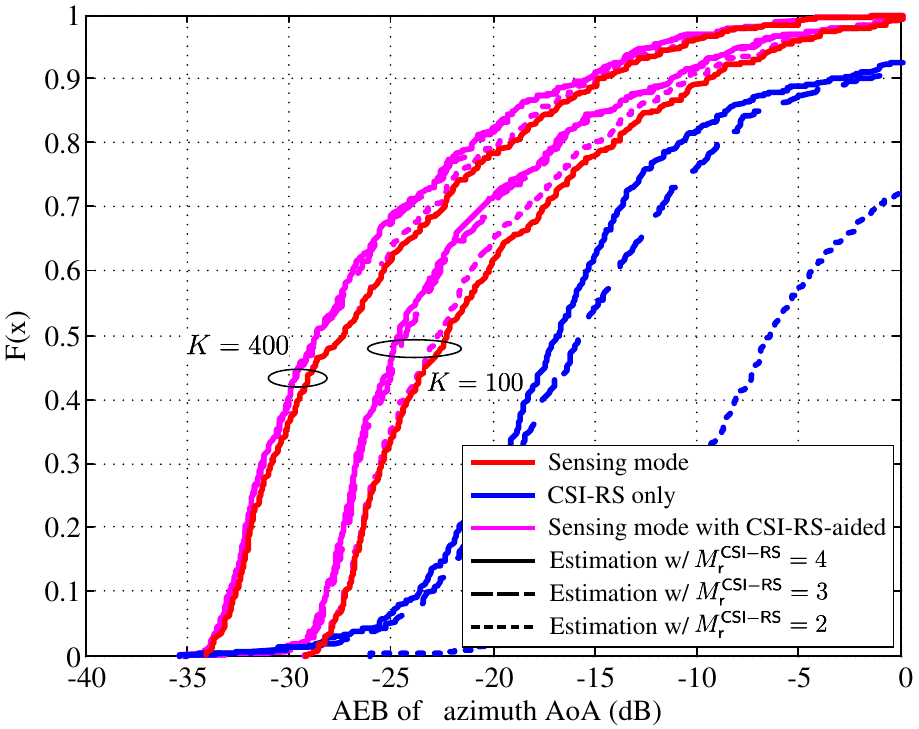}}%
        \caption{CDF of the AEB in $K = 100$ and $K = 400$ with the LoS under medium SNR conditions and $M_{\sf r}^{(\sf se)} = 2$. }\label{fig:track1}
    \end{center}
\end{figure}

We study the effectiveness of CSI-RS-aided environment sensing mentioned in Section III.B. In 5G NR, the UE can estimate the channel parameters using the CSI-RS and calculate the AEB for AOA, represented by $\theta_{{\sf r}, k, l}$ in \eqref{eq:av_channelParms}. We assume that the CSI-RS fixes the transmit beam on the serving beam by 5G NR beam management. The received antenna panel and corresponding Rx beam are determined in the same way as the sensing mode. The number of Rx beams used for channel measurements by CSI-RS is defined as $M_{\sf r}^{\sf CSI-RS}$. The measurements based on the CSI-RS are performed with $N_s^{{\sf CSI-RS}} = 330$ subcarriers. We consider three scenarios: 1) Sensing mode: estimation from accumulated sensing instances $K$ SSBs only, 2) CSI-RS only: estimation from the sensing instance $K$ CSI-RS only, and 3) Sensing mode with CSI-RS-aided: a combination of both estimations. The results are shown in the form of cumulative distribution function (CDF) of AEB for AoA under medium SNR conditions. Fig. \ref{fig:track1} shows the performance of azimuth AoA when the number of sensing instances $K = 100$ and $K = 400$, with the number of receive beams used for receiving $M_{\sf r}^{(\sf se)} = 2$.

The results in Fig. \ref{fig:track1} show the performance of the sensing mode is worse than the estimation from only using CSI-RS in some cases, especially when the sensing instances are insufficient (i.e., $K = 100$ or less). The reason is that the CSI-RS in 5G NR only provides information about the currently received signal and the limitations of the ULA on the mobile phone may inaccurately receive signals with large elevation angles of arrival. On the contrary, the result of sensing mode combines information from multiple orientations over time. Thus, the results suggest a tradeoff between spending time on recent measurements or using many previous measurements to gain situational awareness immediately. However, the combination approach, sensing mode with CSI-RS-aided, always results in better angular accuracy than the other cases. The accuracy gap between the combination approach and the result of the sensing mode can be reduced by increasing the number of receive beams and the amount of time spent. Therefore, the benefit of situational awareness only appears when the accumulation of time $K$ is sufficient.

\subsection{Performance of Beam Foreseeing}

\begin{figure*}
    \begin{center}
        \resizebox{6.65in}{!}{%
            \includegraphics*{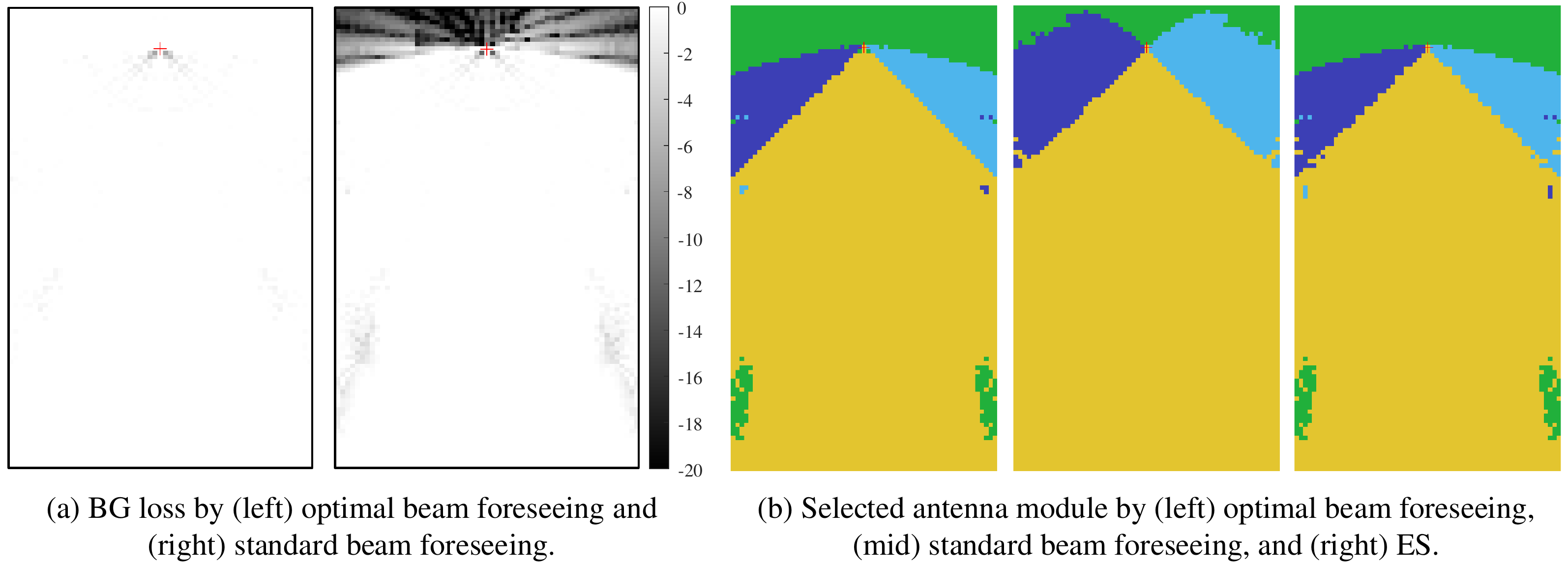}}%
        \caption{Performance of the optimal and standard beam foreseeing approaches.}\label{fig:vsStanda}
    \end{center}
\end{figure*}

\begin{figure*}
    \begin{center}
        \resizebox{6.65in}{!}{%
            \includegraphics*{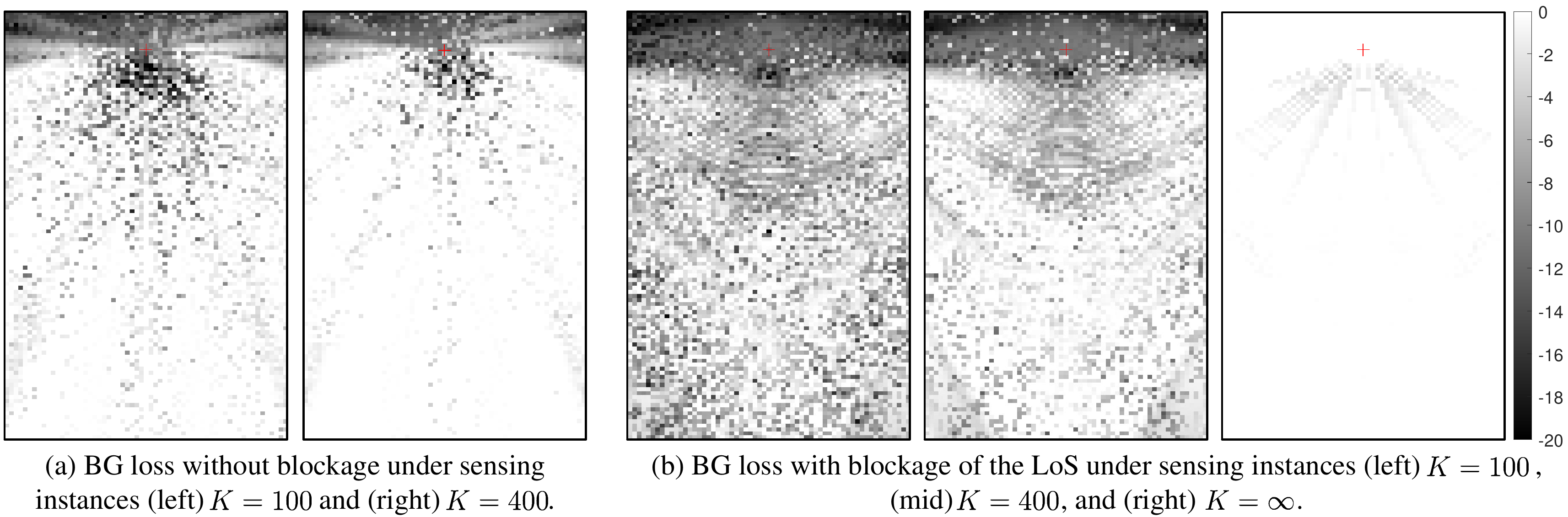}}%
        \caption{Loss of BG for each position of UE on the map (a) without blockage and (b) with blockage under different sensing instances $K$.}\label{fig:overall}
    \end{center}
\end{figure*}

Beam foreseeing is a technique that uses location information to identify the optimal beam pair for communication when the current pair is ineffective. There are two versions of beam foreseeing: standard and optimal. The standard version sets the beam pair based on the LoS path during initial beam and tracks it as the beam pair during subsequent adjustments. This method only utilizes the location information of the BS. In contrast, the optimal version leverages the location information of the BS and vBSs obtained from the sensing mode and uses CSI-RS during beam adjustment to identify the best beam pair from the detected candidates. The number of Tx beams during beam adjustment is limited to $M_{\sf t}^{(\sf tr)} \leq 6$. We use exhaustive scanning (ES) as a benchmark to compare the performance of both methods, which involves scanning all angles and antenna panels. The difference between the BG obtained by beam foreseeing and ES is represented as BG loss.
We calculate the BG loss for different positions of the UE at a height of 1.2 m with a fixed orientation $\qalpha_{{\sf ue},K} = [1, 0, 0, 0]$ in the environment depicted in Fig. \ref{fig:environment}. The variances of the AoAs and AoDs are calculated from \eqref{eq:Theo2} and used to simulate the error from the location information for all $\qv_l$s.

We first assume the AoAs and AoDs are known (i.e., $K = \infty$) without noise to represent an upper bound of beam foreseeing. Fig. \ref{fig:vsStanda}(a) shows the BG loss by different versions. The result obtained using the optimal version is nearly identical to those obtained using ES, with a loss of at most 7 dB. This finding indicates the proposed beam foreseeing method's performance can reach the optimal solution and reduce complexity by nearly 300 times compared to ES. The results of the standard version indicate the BG loss increases as the UE approaches the wall, particularly at the back of the BS. This finding suggests the BG from the LoS path may be smaller than its virtual source (NLoS) due to limitations of distance, beamwidth, or antenna pattern. Moreover, the results of the optimal version in Fig. \ref{fig:vsStanda}(a) demonstrate good estimation results can still be obtained in areas that were not covered by the UE's position in the sensing mode (i.e., outside the gray area of Fig. \ref{fig:environment}).

In Fig. \ref{fig:vsStanda}(b), we show the results of the selected antenna panel by different methods, represented by yellow, light blue, green, and dark blue for panels 1 to 4, respectively. The comparison of antenna panel selections indicates a little difference between the optimal beam foreseeing and ES. In contrast, the standard version, which uses power detection, uses different antenna panels in areas of poor performance. This finding suggests the selection of antenna modules and beams for optimal beam foreseeing is similar to that of ES, resulting in the best performance with extremely low complexity.

Next, we evaluate the performance of beam foreseeing under various conditions, considering errors caused by insufficient $K$ and blockage.
In Fig. \ref{fig:overall}, we evaluate beam foreseeing under poor SNR conditions with $M_{\sf r}^{(\sf se)} = 2$. All the observations found previously can explain the phenomena observed in the figures.
First, Fig. \ref{fig:overall} shows roughly symmetrical results because the vBSs on both sides are symmetrical.
Second, we discuss the impact of angle estimation errors in the absence of blockage. As shown in Fig. \ref{fig:overall}(a), the BG loss increases with the angle estimation error. According to the results of Fig. \ref{fig:vsStanda}(b), antenna panel 3 would be selected for a position at the back of the BS, and the reflection path from $\qv_3$ would be used as the beam pair. However, as mentioned in Section V.A, the AEB obtained by $\qv_3$ cannot obtain sufficiently correct angle estimation. Therefore, it is clear from Fig. \ref{fig:overall}(a) that using $\qv_3$ as the result of beam selection results in extremely poor performance. Moreover, with the increase of sensing instances, most of the areas that can effectively reduce the BG loss are those far from the BS. This is because the closer the UE is to the BS, the more accurate the angle estimation required, and a little deviation in the result of beam selection would lead to a relatively large loss.
Lastly, we discuss the impact of the angle estimation error after the blockage of the LoS. The left figure of Fig. \ref{fig:overall}(b) shows a clear boomerang-shaped pattern resulting from the finite resolutions of beamforming vectors and high angular estimation error. The boomerang-shaped pattern disappears in the right figure of Fig. \ref{fig:overall}(b) when the estimation error decreases to zero. In the right side of Fig. \ref{fig:overall}(b), we observe that the loss appears below the BS, with the maximum loss being approximately 5 dB. This is because the Tx beam that can transmit the signal to the UE located directly below the BS is beyond the range that the BS antenna array can support (from $20^{\circ}$ to $160^{\circ}$).

\section{Conclusion}
This study focused on evaluating the capabilities of 5G NR beam management in providing situational awareness. We assessed the positional accuracy of physical BSs and vBSs, which are critical for environment characterization and identification of potential beam pairs. These assessments were instrumental in predicting the beamforming gain of candidate beams. Furthermore, we developed a beam foreseeing strategy to optimize beam alignment for various scenarios, such as mmWave communication's susceptibility to blockage or random movement. Our analysis highlighted the optimal choices for beam pairs and antenna panels. Simulation results demonstrated that beam foreseeing can rapidly switch beams based on situational awareness, achieving performance close to an ideal scenario.

\section*{Appendix A}

To gain useful insights, we assume the absence of path overlap in the resolvable angle and delay domain.
The FIM in \eqref{eq:channel_FIM} can be represented by submatrices 
$ 
\qF_{\qpsi_{k}}= \blkdiag\left(\qF_{\qpsi_{k, 0}}, \ldots, \qF_{\qpsi_{k, L-1}} \right)
$
with each path being independent.
Each diagonal submatrix of $\qF_{\qpsi_{k}}$ is consisted of a $7 \times 7$ matrix of $\qF_{\qpsi_{k, l}}$ with the elements in $\qF_{\qpsi_{k, l}}$ indicating the parameters of $\qpsi_{k} = \left[\theta_{{\sf r}, k, l}, \varphi_{{\sf r}, k, l}, \theta_{{\sf t}, k, l}, \varphi_{{\sf t}, k, l},  \tau_{k, l}, g_{k, l}^{(\rm R)}, g_{k, l}^{(\rm I)}\right]^{\rm T}$ used for calculation in \eqref{eq:channel_FIM}.

Notably, a common property of antenna characteristics is an approximately symmetric beam pattern, i.e., the beam pattern can be described by an even function. Due to the property of even functions exhibiting an odd function as a derivative, orthogonality between the function and its derivative is obtained. The function of delay has similar characteristics. As a result, each submatrix in $\qF_{\qpsi_{k}}$ is approximately block-diagonal and can be given by
\begin{equation}
\begin{aligned}
&\left[\qF_{\qpsi_{k, l}} \right]_{1:2, 1:2} =\begin{bmatrix}
F_{\theta_{{\sf r}, k, l}, \theta_{{\sf r}, k, l}} & F_{\theta_{{\sf r}, k, l}, \varphi_{{\sf r}, k, l}}\\
F_{\varphi_{{\sf r}, k, l}, \theta_{{\sf r}, k, l}} & F_{\varphi_{{\sf r}, k, l}, \varphi_{{\sf r}, k, l}}\end{bmatrix},\\
&\left[\qF_{\qpsi_{k, l}} \right]_{3:4, 3:4} =\begin{bmatrix}
F_{\theta_{{\sf t}, k, l}, \theta_{{\sf t}, k, l}} & F_{\theta_{{\sf t}, k, l}, \varphi_{{\sf t}, k, l}}\\
F_{\varphi_{{\sf t}, k, l}, \theta_{{\sf t}, k, l}} & F_{\varphi_{{\sf t}, k, l}, \varphi_{{\sf t}, k, l}}\end{bmatrix},\\
&\left[\qF_{\qpsi_{k, l}} \right]_{5, 5} = F_{\tau_{k, l}, \tau_{k, l}},\\
&\left[\qF_{\qpsi_{k, l}} \right]_{6, 6} = F_{g_{k, l}^{(\rm R)}, g_{k, l}^{(\rm R)}},\\
&\left[\qF_{\qpsi_{k, l}} \right]_{7, 7} = F_{g_{k, l}^{(\rm I)}, g_{k, l}^{(\rm I)}}.
\end{aligned}
\end{equation}

To focus on the channel parameters related to partial localization, we derive the EFIM using the Schur complement. Finally, we obtain the EFIM for the available channel parameters as
\begin{equation}\label{eq:EFIM_feat}
\qF_{\dot{\qpsi}_{k, l}}= \blkdiag\left(\left[\qF_{\qpsi_{k, l}} \right]_{1:2, 1:2}, \left[\qF_{\qpsi_{k, l}} \right]_{5, 5}\right).
\end{equation}

\begin{figure*}[!t]
\begin{subequations}\label{eq:T_vele}
\begin{align}
\frac{\partial \theta_{{\sf r}, k, l}}{\partial \Delta\qalpha} &= \frac{1}{(v'_{{\sf x}, l, k})^2+(v'_{{\sf y}, l, k})^2}\left(v'_{{\sf x}, l, k}\frac{\partial \qr_{{\sf ue},k,2}^{\rm T}}{\partial \Delta\qalpha}-v'_{{\sf y}, l, k}\frac{\partial \qr_{{\sf ue},k,1}^{\rm T}}{\partial \Delta\qalpha} \right)(\qv_{l}-\qp_k) \\
\frac{\partial \varphi_{{\sf r}, k, l}}{\partial \Delta\qalpha} &= \frac{-1}{\sqrt{(v'_{{\sf x}, l, k})^2+(v'_{{\sf y}, l, k})^2}}\frac{\partial \qr_{{\sf ue},k,3}^{\rm T}}{\partial \Delta\qalpha}(\qv_{l}-\qp_k)\\
\frac{\partial \theta_{{\sf r}, k, l}}{\partial \qv_{l}} &= \frac{1}{(v'_{{\sf x}, l, k})^2+(v'_{{\sf y}, l, k})^2}\left(v'_{{\sf x}, l, k}\qr_{{\sf ue},k,2}-v'_{{\sf y}, l, k} \qr_{{\sf ue},k,1} \right)\\
\frac{\partial \varphi_{{\sf r}, k, l}}{\partial \qv_{l}} &= \frac{-1}{\sqrt{(v'_{{\sf x}, l, k})^2+(v'_{{\sf y}, l, k})^2}} \left( \qr_{{\sf ue},k,3}-\frac{v'_{{\sf z}, l, k}}{\|\qv'_{l, k}\|^2}(\qv_{l}-\qp_k)\right) \\
\frac{\partial \theta_{{\sf t}, k, l}}{\partial \qv_{l}}&=\left\lbrace \begin{array}{ll}
\frac{1}{(p'_{{\sf x}, l, k})^2+(p'_{{\sf y}, l, k})^2}\left(p'_{{\sf y}, l, k} \qr_{{\sf bs},1}-p'_{{\sf x}, l, k}\qr_{{\sf bs},2} \right), & \text{if $\qv_{l}$ is not mirrored horizontally}\\
\begin{aligned}
\frac{-1}{(p'_{{\sf x}, l, k})^2+(p'_{{\sf y}, l, k})^2}&\left(p'_{{\sf y}, l, k} \qr_{{\sf bs},1}-p'_{{\sf x}, l, k}\qr_{{\sf bs},2} \right) \\+\frac{2}{(v'_{{\sf x}, l})^2+(v'_{{\sf y}, l})^2}&\left(v'_{{\sf y}, l} \qr_{{\sf bs},1}+v'_{{\sf x}, l}\qr_{{\sf bs},2} \right),
\end{aligned}
 & \text{if $\qv_{l}$ is mirrored horizontally}
\end{array}\right.\\
\frac{\partial \varphi_{{\sf t}, k, l}}{\partial \qv_{l}}&=\left\lbrace \begin{array}{ll}
\frac{-1}{\sqrt{(p'_{{\sf x}, l, k})^2+(p'_{{\sf y}, l, k})^2}} \left( \frac{p'_{{\sf z}, l, k}}{\|\qp'_{l, k}\|^2}(\qp_k-\qv_{l})-\qr_{{\sf bs},3}\right),& \text{if $\qv_{l}$ is not mirrored vertically}\\
\begin{aligned}
\frac{1}{\sqrt{(p'_{{\sf x}, l, k})^2+(p'_{{\sf y}, l, k})^2}} &\left( \frac{p'_{{\sf z}, l, k}}{\|\qp'_{l, k}\|^2}(\qp_k-\qv_{l})-\qr_{{\sf bs},3}\right)\\+\frac{2}{\sqrt{(v'_{{\sf x}, l})^2+(v'_{{\sf y}, l})^2}} &\left( \frac{v'_{{\sf z}, l}}{\|\qv'_{l}\|^2}(\qv_{l}-\qv_{0})-\qr_{{\sf bs},3}\right) ,
\end{aligned}& \text{if $\qv_{l}$ is mirrored vertically}
\end{array}\right.\\
\frac{\partial \theta_{{\sf t}, k, l}}{\partial \qv_{0}}&=\frac{-2}{(v'_{{\sf x}, l})^2+(v'_{{\sf y}, l})^2}\left(v'_{{\sf y}, l} \qr_{{\sf bs},1}+v'_{{\sf x}, l}\qr_{{\sf bs},2} \right),~~~~~~~~~~\text{if $\qv_{l}$ is mirrored horizontally}\\
\frac{\partial \varphi_{{\sf t}, k, l}}{\partial \qv_{0}}&=\frac{-2}{\sqrt{(v'_{{\sf x}, l})^2+(v'_{{\sf y}, l})^2}} \left( \frac{v'_{{\sf z}, l}}{\|\qv'_{l}\|^2}(\qv_{l}-\qv_{0})-\qr_{{\sf bs},3}\right),\text{if $\qv_{l}$ is mirrored vertically}
\end{align}
\end{subequations}
\hrulefill
\vspace*{4pt}
\end{figure*}

\section*{Appendix B}
We determine the unconstrained FIM for spatial parameters $\underline{\qphi}$ by transforming from channel parameters $\dot\qpsi_k$. We define $\qR(\qalpha_{{\sf ue},k}) = [\qr_{{\sf ue},k,1}|\qr_{{\sf ue},k,2}|\qr_{{\sf ue},k,3}]$ and $\qR(\qalpha_{{\sf bs}}) = [\qr_{{\sf bs},1}|\qr_{{\sf bs},2}|\qr_{{\sf bs},3}]$ as the column vectors of the rotation matrix for UE and BS. Using \eqref{eq:Qua_delta}, the relationship between the UE and BS rotation matrices can be written as $\qR(\qalpha_{{\sf bs}}) = \qR(\Delta\qalpha)\qR(\qalpha_{{\sf ue},0})$. If the UE's relative orientation $\qR_{k}$ at different times can be obtained from VIO, then the UE's rotation matrix at the $k$-th time can be written as $\qR(\qalpha_{{\sf ue},k}) = \qR_{k}\qR(\qalpha_{{\sf ue},0})$.

Using the geometric transformation in \eqref{eq:geo}, we transform the channel parameters into the spatial parameters. The elements in the transformation matrix in \eqref{eq:T_psi} can be expressed as \eqref{eq:T_vele} top on the next page and the remaining elements are zero.
We calculate the derivatives of the relative orientation vector $\Delta\qalpha\triangleq[\Delta\alpha_0, \Delta\alpha_1, \Delta\alpha_2, \Delta\alpha_3]$ as
\begin{equation}
\frac{\partial \qr_{{\sf ue},k,1}^{\rm T}}{\partial \Delta\alpha_0} = \qr_{{\sf bs},1}^{\rm T}\frac{\partial \qR(\Delta\qalpha)}{\partial \Delta\alpha_0}\qR_{k}^{\rm T},
\end{equation}
and similarly for other rotation vectors and relative orientation elements.


\section*{Appendix C}

The OEB is defined as the angle error between the true and estimated orientation vectors, represented through quaternions denoted by $\qalpha = [\alpha_0, \alpha_1, \alpha_2, \alpha_3]$ and $\hat\qalpha = [\hat\alpha_0, \hat\alpha_1, \hat\alpha_2, \hat\alpha_3]$.
The error from $\hat\qalpha$ to $\qalpha$ can be calculated by quaternion multiplicative \cite{Ori} error as
$
    \delta\qalpha = {\rm Q_m}(\qalpha, \hat\qalpha^{-1}) = [\delta\alpha_0, \delta\alpha_1, \delta\alpha_2, \delta\alpha_3],
$
where $\delta\qalpha$ represents the shortest rotation between true and estimated orientation and $\hat\qalpha^{-1}$ represents the reciprocal of $\hat\qalpha$. For a unit quaternion, the reciprocal is the same as conjugate \cite{QuaCbook}, written as $\hat\qalpha^{-1} = [\hat\alpha_0, -\hat\alpha_1, -\hat\alpha_2, -\hat\alpha_3]$. With the definition of quaternion multiplication, we obtain $\delta\alpha_0 = \alpha_0 \hat\alpha_0+\alpha_1 \hat\alpha_1+\alpha_2 \hat\alpha_2+\alpha_3 \hat\alpha_3$. Since the quaternion can be expressed as a rotation around a unit orientation vector, the angle error is calculated as $2 \cos^{-1}(\delta\alpha_0)$ \cite{QuaCbook}.
The error bound obtained from constrained FIM is
$
\mathrm{tr}\{[ \qU ( \tilde{\qF}_{\underline{\qphi}}) ^{-1}\qU^{\rm T}] _{1:4, 1:4} \}
=(\alpha_0 - \hat\alpha_0)^2 + (\alpha_1 - \hat\alpha_1)^2+ (\alpha_2 - \hat\alpha_2)^2 + (\alpha_3 - \hat\alpha_3)^2
= 2 - 2 \delta\alpha_0,
$
Consequently, we obtain $\delta\alpha_0 = 1-\mathrm{tr}\{[ \qU ( \tilde{\qF}_{\underline{\qphi}}) ^{-1}\qU^{\rm T}] _{1:4, 1:4} \}/2$ and the OEB as \eqref{eq:EB_Dalpha}.

{\renewcommand{\baselinestretch}{1.1}
    \begin{footnotesize}
        \bibliographystyle{IEEEtran}
        \bibliography{IEEEabrv,References}
\end{footnotesize}}

\end{document}